\documentclass[%
    a4paper,
    nofootinbib,
    preprintnumbers,
    notitlepage,
    aps,
    prd
]{revtex4-2}

\usepackage{geometry} 
\usepackage[utf8]{inputenc} 
\usepackage[T1]{fontenc} 
\usepackage{lmodern}
\usepackage[british]{babel}
\usepackage{graphicx}
\graphicspath{{figures/}} 
\usepackage[pdfusetitle]{hyperref}

\usepackage{microtype} 
\usepackage{xfrac} 
\makeatletter
\g@addto@macro\@floatboxreset\centering
\makeatother

\usepackage[intlimits]{amsmath} 
\usepackage{amssymb} 
\usepackage{bm} 
\usepackage{bbm} 
\numberwithin{equation}{section} 
\allowdisplaybreaks 

\usepackage[capitalise,nameinlink]{cleveref} 

\usepackage{booktabs} 
\usepackage{array} 
\newcolumntype{L}{>{$}l<{$}} 
\usepackage{multirow}

\usepackage[italic]{hepnames} 

\usepackage{braket} 
\usepackage{slashed} 
\usepackage{siunitx} 
\DeclareSIUnit{\barn}{b}
\DeclareSIUnit{\fb}{\femto\barn}
\DeclareSIUnit{\invps}{\per\ps}
\usepackage{suffix} 

\usepackage{enumitem}
\newlist{properties}{enumerate}{1}
\setlist[properties,1]{label=\textbf{P\arabic*},ref=\textbf{P\arabic*}}
\crefname{propertiesi}{property}{properties}

\newcommand{\ie}{\textit{i.e.}\xspace}
\newcommand{\eg}{\textit{e.g.}\xspace}

\newcommand{\Kl}[1]{\ensuremath{K_{l#1}}\xspace}
\newcommand{\Kmutwo}{\ensuremath{K_{\mu 2}}\xspace}
\newcommand{\KS}{\ensuremath{K^0_\text{S}}}
\newcommand{\KL}{\ensuremath{K^0_\text{L}}}
\newcommand{\Omnes}{Omn\`{e}s\xspace}
\newcommand{\V}[1]{\ensuremath{V_{#1}^{}}\xspace} 
\WithSuffix\newcommand\V*[1]{\ensuremath{V_{#1}^*}\xspace}
\newcommand{\absV}[1]{\ensuremath{|V_{#1}^{}|}\xspace}
\newcommand{\Ci}[2][\phantom{x}]{\ensuremath{C_{#2}^{#1}}}
\newcommand{\CVL}[1][\phantom{x}]{\Ci[#1]{V,L}\xspace}
\newcommand{\CVR}[1][\phantom{x}]{\Ci[#1]{V,R}\xspace}

\newcommand{\code}[1]{\texttt{#1}\xspace}
\newcommand{\EOS}{\code{EOS}}
\newcommand{\dynesty}{\code{dynesty}}

\newcommand{\vecx}{\vec{x}\xspace}

\newcommand{\tp}{\ensuremath{t_+}\xspace}
\newcommand{\tm}{\ensuremath{t_-}\xspace}
\newcommand{\bp}[1]{\ensuremath{b_{#1}^{(+)}}\xspace}
\newcommand{\bz}[1]{\ensuremath{b_{#1}^{(0)}}\xspace}

\newcommand{\Kresonance}[1]{\ensuremath{K^*(#1)}\xspace}
\newcommand{\Kzeroresonance}[1]{\ensuremath{K^*_0(#1)}\xspace}

\begin{document}

\author{Matthew Kirk}
\affiliation{%
    Institute for Particle Physics Phenomenology and Department of Physics, Durham University, Durham DH1 3LE, UK
}
\author{Danny van Dyk}
\affiliation{%
    Institute for Particle Physics Phenomenology and Department of Physics, Durham University, Durham DH1 3LE, UK
}
\title{A Global Determination of \absV{us}}
\preprint{EOS-2025-05, IPPP/25/75}

\begin{abstract}
    In light of ongoing issues with the first-row unitarity test of the CKM matrix,
    we showcase a global fit to measurements of $K\to \ell \nu$, $K\to \pi\ell \nu$,
    $\tau \to K\nu$, and $\tau\to K\pi \nu$ decays for the first time.
    Fitting the semileptonic and 3-body $\tau$ decay data simultaneously becomes computationally feasible
    because we employ a simple form factor parametrisation for the $K\pi$ form factor that manifestly connects
    the semileptonic and pair-production regions.
    We find good agreement with the data and use our analyses to infer $|V_{us}|$ and the parameters
    for the $K\pi$ form factors.
    Our result for $|V_{us}|$ sits close to the results extracted from exclusive $K_{\ell 2}$ and somewhat above 
    the $K_{\ell 3}$ decay and inclusive $\tau$ decay result.
    Moreover, our results for the $K\to \pi$ vector form factor at zero momentum transfer
    are compatible with lattice QCD determinations, despite not using lattice QCD inputs for this quantity in our fit of the hadronic matrix elements.
    We are further able to determine some of the pole parameters of the individual scalar and vector
    $K\pi$ resonances with masses below the $\tau$ mass.
    We caution that our results account only for short-distance electromagnetic corrections
    but not long-distance contributions; this is
    due to the lack of a consistent description of long-distance corrections to the $K\pi$
    matrix elements both above and below threshold for an arbitrary model of the form factors.
\end{abstract}

\maketitle

\section{Introduction}
\label{sec:intro}

Since the discovery of the Higgs Boson, the complete field content of the Standard Model (SM) has now been discovered, and so far there has been no direct signal of any new particles at the LHC.
As such, we increasingly turn to precision tests of the SM in the hope of discovering a significant deviation.
A clear prediction of the SM is the unitarity of the CKM matrix, manifesting as a wealth of predicted relationships between elements of that matrix which can be tested against experimental determinations.
Amongst others, we have the ``first-row unitarity'' prediction:
\begin{equation}
    |\V{ud}|^2 + |\V{us}|^2 + |\V{ub}|^2 = 1 \,.
\end{equation}
For a long time, this relationship held within experimental and theoretical uncertainties, but increasing precision on both fronts means we can test this relation at unprecedented level.
Taking for example the PDG averages for \V{ud}%
\footnote{%
    We note several recent new calculations of various corrections to $\beta$
    decay~\cite{Gorchtein:2025wli,Moretti:2025qxt,Cao:2025zxs,Crosas:2025xyv}
    that could shift the value of \V{ud}. We continue to use the PDG value for
    comparison and look forward to a comprehensive update including the latest calculations.
} and \V{us} \cite{ParticleDataGroup:2024cfk,PDG:Vud_Vus_review} and neglecting the contribution due to \V{ub} on account of its size,
one obtains
\begin{equation}
\label{eq:first-row-unitarity-PDG}
    |\V{ud}|^2 + |\V{us}|^2 = 0.9983 \pm 0.0006 (\V{ud}) \pm 0.0004 (\V{us}) \,,
\end{equation}
showing a discrepancy with respect to unitarity of more than \qty{2}{\sigma}.
With uncertainties at the level of less than one part per mille, this is currently the most stringent test of unitarity in
the CKM matrix.
Using a single value for \absV{us}, as we have done here, hides another discrepancy, whereby \absV{us} values extracted from leptonic kaon decays ($K \to \mu \nu$, commonly known as \Kmutwo%
\footnote{%
    In fact there is yet another layer, in that the value of \V{us} from \Kmutwo is typically obtained from the experimental
    ratio $\Gamma(K \to \mu \nu) / \Gamma(\pi \to \mu \nu)$, a lattice determination of $f_K / f_\pi$, and then either using
    CKM unitarity to eliminate \V{ud}, or using a \V{ud} from e.g.\ super-allowed $\beta$ decays.
}%
) disagree with those from semi-leptonic kaon decays ($K \to \pi \ell \nu$, commonly known as \Kl3), leading to the PDG result above receiving an inflated error.
Additional determinations of \absV{us} from inclusive \Ptau decays are less precise than either of these methods, but appear to give yet another discrepant result.%

The tension in this ``first-row unitarity'' test has been interpreted in terms of a variety of Beyond the Standard Model (BSM) physics.
Of particular interest for BSM model building is the observed discrepancy between the \absV{us} extractions from two body and three body kaon decays, as this could indicate a non-zero right handed current -- this thread has been followed in Refs.~\cite{Grossman:2019bzp,Belfatto:2021jhf,Crivellin:2021njn,Crivellin:2021bkd,Crivellin:2022rhw,Cirigliano:2022yyo,Cirigliano:2023nol,Belfatto:2023tbv}.
\\

In this work we conduct a pilot study that simultaneously analyses data from \Kl2, \Kl3, $\tau \to K\pi \nu$ and $\tau \to K \nu$ decays, which are all $s \to u \ell \nu$ transitions (and in the SM all governed by a single effective operator).
At leading order in $\alpha_e$, the three-body kaon and \Ptau decays are jointly described by two hadronic form factors, evaluated in different regions of phase space.
We are able to conduct a global analysis at leading-order in $\alpha_e$ by making use of a new novel parameterisation of these form factors,
which explicitly takes into account the resonances present in the $\tau \to K \pi \nu$ decay spectrum (shown in \cref{fig:res:tau-decay}),
as first developed in Ref.~\cite{Kirk:2024oyl}.
Beyond leading order in $\alpha_e$, long-distance process-specific corrections arise.
These have been calculated for \Kl3 decays in Ref.~\cite{Cirigliano:2008wn}, while for two hadron \Ptau decays
only partial results are available: corrections assuming point-like mesons are provided in Refs.~\cite{Antonelli:2013usa,Flores-Baez:2013eba};
meson structure dependent corrections with a real emitted photon have been recently calculated in Ref.~\cite{Escribano:2023seb};
there has not yet been a calculation of the structure-dependent virtual photon corrections, however.
Furthermore, these corrections are not consistently available for a common form factor parametrisation that covers both the semileptonic and pair-production phase space regimes.
Hence, they are not included in our analysis.
To go beyond this pilot study, both the process-specific electromagnetic corrections and isospin-breaking effects must be accounted for.

\section{Theoretical Framework}

Weak decay processes are very commonly described within the framework of the Weak Effective Theory (WET), which
is constructed from SM fields with masses below the scale of electro-weak symmetry breaking.
Within this theory, one identifies sectors corresponding to sets of operators that do not mix under the renormalisation 
group equations at leading order in the Fermi constant $G_\text{F} \sim g^2/M_W^2$~\cite{Aebischer:2017ugx}.
For the kaon and \Ptau decays at hand, the relevant sector is described by the Lagrangian
\begin{equation}
    \label{eq:intro:wet:lagrangian}
    \mathcal{L}^{us\ell\nu}
        = \frac{4 G_\text{F}}{\sqrt{2}} \V{us} \sum_i \Ci[\ell]{i}(\mu) \mathcal{O}_i^\ell + \text{h.c.}\,.
\end{equation}
Here $\mathcal{O}_i^\ell$ are a set of local operators of mass dimension six involving an $\ell$-flavoured
charged lepton field, and $\Ci[\ell]{i}(\mu)$ are their WET Wilson coefficients.
In the SM, only a single operator with $V-A$ structure contributes, \ie, only a single
operator features a non-vanishing Wilson coefficient.
The operator reads
\begin{equation}
    \label{eq:intro:wet:OVL}
    \mathcal{O}_{V,L}^\ell = [\bar{u}\gamma^\mu P_L s]\, [\bar{\ell} \gamma_\mu P_L \nu]\,,
\end{equation}
where $2 P_{L/R} \equiv 1 \mp \gamma_5$. To order $\alpha_e$, the SM value of its Wilson coefficient
is lepton-flavour universal and reads
\begin{equation}
    \label{eq:intro:wet:CVL-SM}
    \CVL[\ell](\mu)
        = 1 + \frac{\alpha_e}{\pi} \ln \left(\frac{M_Z}{\mu}\right) \simeq 1.01\,.
\end{equation}
The numerical approximation above holds for $\mu \simeq \qty{1}{\GeV}$,
which is chosen close to the geometric mean of $M_K$ and $M_\tau$ and
therefore minimizes potential logarithmic enhancements in the matrix elements of the operators.
We assume $\mu = \qty{1}{\GeV}$ whenever we discuss values of WET Wilson coefficients throughout this work.\\

Allowing for Beyond the Standard Model (BSM) effects, but still assuming the presence of only
left-handed neutrinos, there exist four more operators that can contribute to the decays under consideration
at mass dimension six. For the purposes of this study we briefly consider only contributions by a
right-handed $\bar{u}s$ current, corresponding to the operator
\begin{equation}
    \label{eq:intro:wet:OVR}
    \mathcal{O}_{V,R}^\ell = [\bar{u}\gamma^\mu P_R s]\, [\bar{\ell} \gamma_\mu P_L \nu]\,.
\end{equation}

\section{Hadronic Matrix Elements}
\label{sec:hmes}

In order to calculate amplitudes and (subsequently) decay rates and other observables based
on the Lagrangian \cref{eq:intro:wet:lagrangian},
one needs access to the (hadronic) matrix element of the $u s \ell \nu$ operators for the
appropriate external hadronic states.
In this section, we discuss the hadronic matrix elements relevant to this study.

\subsection{Preliminaries}
\label{sec:hmes:preliminaries}

To determine \absV{us} from individual processes, information about process-specific hadronic matrix elements are required.
In the description of leptonic decays of a kaon $K\to \ell\nu$ (dubbed \Kl2),
to order $\alpha_e^0$, only a single decay constant $f_K$ emerges.
The latter is commonly defined as~\cite{FlavourLatticeAveragingGroupFLAG:2024oxs}
\begin{equation}
    \braket{0 | \bar{u} \gamma^\mu \gamma_5 s | K^-(p_K)}
        \equiv
            -i p_K^\mu f_K\,.
\end{equation}

In the description of the semileptonic decay of a kaon $K\to \pi \ell\nu$ (dubbed \Kl3),
two hadronic form factors $f_+$ and $f_0$ emerge at order $\alpha_e^0$.
These form factors are scalar-valued functions of the dilepton mass square $q^2$.
They are commonly defined as\footnote{%
    We define the form factors in terms of the $\bar{K}^0$ initial state, since experimental data for the
    semileptonic rate of a neutral kaon are much more precise than for the charged kaon.
    Assuming isospin symmetry, the form factors for $K^0_{L,S} \to \pi^+$ and $K^- \to \pi^0$ transitions are related to the $\bar{K}^0 \to \pi^+$ one in the following way: $\braket{\pi^+ | j_{\bar{u} s} | \bar{K}^0} =  \pm  \sqrt{2} \braket{\pi^+ | j_{\bar{u} s} | K^0_{L,S}} = -\sqrt{2} \braket{\pi^0 | j_{\bar{u} s} | K^-}$
}
\begin{equation}
    \label{eq:hmes:k-to-pi}
    \braket{\pi^+ (p_\pi) | \bar{u} \gamma_\mu s | \bar{K}^0 (p_K)}
        \equiv
            f_+ (q^2) \left[ (p_K + p_\pi)_\mu - \frac{M_K^2 - M_\pi^2}{q^2} q_\mu \right]
            + f_0(q^2) \left[ \frac{M_K^2 - M_\pi^2}{q^2} q_\mu \right]\,,
\end{equation}
and the relevant phase space is $m_\ell^2 \leq q^2 \leq (M_K - M_\pi)^2 \equiv \tm$.
To avoid an unphysical pole in the hadronic matrix element, the form factors must
coincide at $q^2 = 0$, \ie, they must fulfil $f_+(0) = f_0(0)$.
Since the phase space interval is numerically small, the form factors are commonly expanded
in $q^2$ around the point $q^2 = 0$ to linear or quadratic order, however, more sophisticated
parametrisations are discussed in the literature~\cite{ParticleDataGroup:2024cfk,PDG:FF-review}.
The world average of lattice computations provides a value for
$f_+(0)$, \ie, the normalisation at zero momentum transfer~\cite{FlavourLatticeAveragingGroupFLAG:2024oxs}.
In the definition \cref{eq:hmes:k-to-pi}, $f_+$ parametrizes a $P$-wave $K\pi$ state
and $f_0$ parametrizes an $S$-wave $K\pi$ state.
The form factor $f_0$ also emerges in hadronic matrix elements of the scalar $\bar{u}s$ current,
\begin{equation}
    \braket{\pi^+ (p_\pi) | \bar{u} s | \bar{K}^0 (p_K)}
        \equiv f_0 (q^2) \frac{M_K^2 - M_\pi^2}{m_s - m_u}\,.
\end{equation}

In the description of the hadronic decay of a $\tau$ to a two-body strange final state $\tau\to K\pi \nu$,
both form factors emerge once more:
the relevant hadronic matrix elements are related by crossing symmetry to the one for the \Kl3
decay. The phase space for the \Ptau decay is $\tp \equiv (M_K + M_\pi)^2 \leq q^2 \leq m_\tau^2$.
However, the point $q^2 = (M_K + M_\pi)^2$ is a branch point of the form factors corresponding to the onset
of the $K\pi$ production threshold. As a consequence, both form factors become multivalued and develop
an imaginary part. In experimental studies, they are commonly parametrized using a sum of Breit-Wigner terms~\cite{Belle:2007goc}.
More sophisticated analyses (involving unitarity and analyticity arguments, and \eg based on an \Omnes representation)
are available~\cite{Jamin:2006tk,Jamin:2008qg,Boito:2008fq,Boito:2010me,Rendon:2021nvu}.\\

When inferring \absV{us} from a single process only, relations between the various hadronic matrix elements are not relevant.
However, for a global determination of \absV{us}, these relations are essential to ensure consistency of the
determination.

The first of these relations emerges when using \Kl3 and \Ptau decays in a joint analysis. Such an analysis
needs to ensure that the two form factors are consistently used across phase space boundaries.
To this end, we adapt a recently proposed parametrisation for the pion form factor~\cite{Kirk:2024oyl} to
both of the $K\pi$ form factors; its discussion is relegated to \cref{sec:hmes:param}.

The second of these relations emerges when using \Kl2 and \Kl3 or \Ptau decays in a joint analysis.
It connects the $f_0$ form factor with the kaon decay constant $f_K$ at two $q^2$ points, by means of the so-called
Callan-Treiman relations~\cite{Callan:1966hu,Dashen:1969bh,Oehme:1966qaf}. These relations emerge from chiral symmetry
arguments and the exploitation of equal-time commutation relations at the zero-recoil point
$q^2 = (m_K^2 - m_\pi^2)$ using the soft-pion theorem~\cite{Callan:1966hu,Dashen:1969bh}
and the point $q^2 = (m_\pi^2 - m_K^2)$ using the soft-kaon theorem~\cite{Oehme:1966qaf}.
These (generalized) Callan-Treiman relations read
\begin{equation}
    f_0 (m_K^2 - m_\pi^2) = \frac{f_{K}}{f_{\pi}} + \Delta_\textrm{CT}
    \,, \quad
    f_0(m_\pi^2 - m_K^2) = \frac{f_{\pi}}{f_{K}} + \Delta_{\widetilde{\textrm{CT}}}\,.
\end{equation}
The corrections were calculated to next-to-leading order in chiral perturbation theory~\cite{Gasser:1984ux}, giving
\begin{equation}
\label{eq:DeltaCT-params}
    \Delta_\textrm{CT} = \num{-3.5 +- 8 e-3} \,, \quad \Delta_{\widetilde{\textrm{CT}}} = \num{0.03 +- 0.03} \,.
\end{equation}
The error on $\Delta_\textrm{CT}$ is directly taken from Ref.~\cite{Bernard:2009zm}\footnote{%
    Ref.~\cite{Bernard:2009zm} in turn cites a private communication with H.~Leutwyler.
}, while the error on $\Delta_{\widetilde{\textrm{CT}}}$ is based on the range $-0.035 < \Delta_{\widetilde{\textrm{CT}}} < 0.11$ provided in Ref.~\cite{Bernard:2009zm}.

\subsection{Properties}
\label{sec:hmes:properties}

\begin{table}[t]
    \centering
    \begin{tabular}{r @{\hskip 2em} c @{\hskip 2em} S[table-format=1.3] S[table-format=1.3] l}
        \toprule
        ~
            & ~
            & \multicolumn{2}{c}{approx.}
            & ~
            \\
        name
            & $S^P$
            & \multicolumn{1}{c}{mass [\unit{\GeV}]}
            & \multicolumn{1}{c}{width [\unit{\GeV}]}
            & channels
            \\
        \midrule
        $K^*(896)$
            & $1^-$
            & 0.890
            & 0.052
            & $K\pi$
            \\
        $K^*(1410)$
            & $1^-$
            & 1.368
            & 0.212
            & $K\pi$, $K\pi\pi$ (via $K^*\pi$, $K\rho$)
            \\
        $K^*(1680)$
            & $1^-$
            & 1.718
            & 0.320
            & $K\pi$, $K\pi\pi$ (via $K^*\pi$, $K\rho$), $K\eta$
            \\
        \midrule
        $\kappa = K_0^*(700)$
            & $0^+$
            & 0.680
            & 0.300
            & $K\pi$
            \\
        $K_0^*(1430)$
            & $0^+$
            & 1.425
            & 0.270
            & $K\pi$, $K\eta$
            \\
        $K_0^*(1950)$
            & $0^+$
            & 1.957
            & 0.170
            & $K\pi$
            \\
        \bottomrule
    \end{tabular}
    \caption{%
        Summary of the properties of the known resonances emerging in the description of the form factors
        $f_+$ and $f_0$, as recorded in the PDG review~\cite{ParticleDataGroup:2024cfk}.
    }
    \label{tab:hmes:resonances}
\end{table}

\begin{figure}[t]
    \centering
    \includegraphics[width=0.45\linewidth]{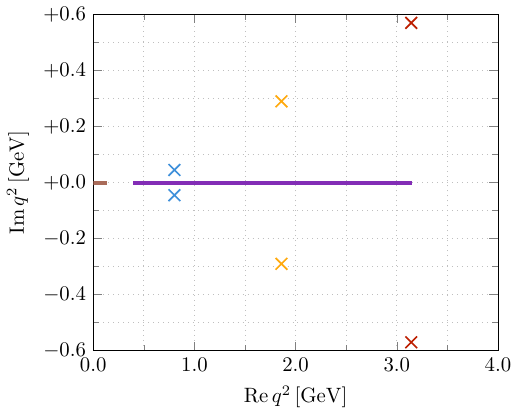}
    \hfill
    \includegraphics[width=0.45\linewidth]{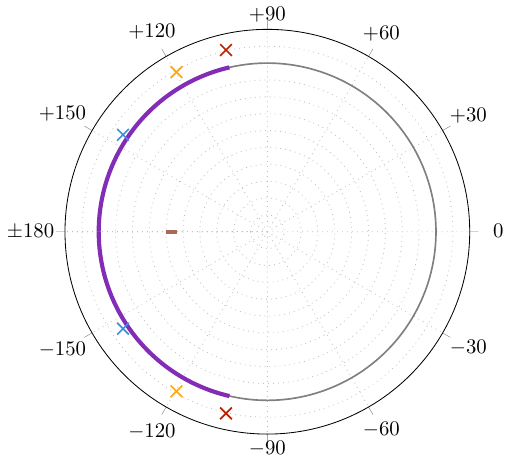}
    \caption{%
        Illustration of the relevant phase space regions and locations of the resonance poles in the complex
        $q^2$ (left) and $z$ (right) planes for the $f_+$ form factor.
        The phase space of $\tau \to K \pi \nu$ is illustrated as purple lines and curves.
        The phase space of $K\to \pi \ell^-\bar\nu$ is illustrated as brown lines.
        The resonance poles are marked with coloured crosses ($K^*(896)$: light blue, $K^*(1410)$: yellow, $K^*(1680)$: red).
    }
    \label{fig:hmes:illustration}
\end{figure}

The form factors $f_+(q^2)$ and $f_0(q^2)$ exhibit useful properties that can be exploited in their parametrisation.
\begin{properties}
    \item \label{prop:hmes:properties:no-bound-states}
    Both form factors parametrize $K\pi$ states with definite spin-parity quantum numbers $J^P$. In the $f_+$ case,
    these are $J^P = 1^-$, while in the $f_0$ case these are $J^P = 0^+$. Since both the kaon and the pion are spin $S^P = 0^-$
    states, the total angular momentum coincides with the orbital angular momentum of the two mesons.
    There are no known states with these quantum numbers below the $K\pi$ threshold. Hence, the form factors are free of
    bound-state poles, which would manifest as poles on the first Riemann sheet.
    
    \item \label{prop:hmes:properties:finite-threshold}
    At the $K \pi$ pair production threshold $\tp$, the form factors are finite.

    \item \label{prop:hmes:properties:angular-momentum-threshold}
    Starting at $t_+$, the two form factors each have a right-handed branch cut and therefore develop an imaginary part.
    For $f_+$ with $L=1$, this discontinuity develops as $(t_+ - q^2)^{3/2}$, while for $f_0$ with $L=0$, it is proportional to $(t_+ - q^2)^{1/2}$.

    \item \label{prop:hmes:properties:further-cuts}
    Further right-handed branch cuts start at higher pair-production thresholds, \eg, $(M_K + 2M_\pi)^2$ and $(M_K + M_\eta)^2$.

    \item \label{prop:hmes:properties:real-below-threshold}
    Using a suitable phase convention, both form factors are real-valued functions on the real $q^2$ axis for
    $q^2 < t_+$. As a consequence, when expanding the form factors around any point on the real $q^2$ axis to the left of $t_+$,
    the expansion coefficients must be real-valued.

    \item \label{prop:hmes:properties:above-threshold-resonances}
    The combination of individual branch cuts show an intricate pattern of resonances, which reflect the presence
    of a series of poles off the real axis on the second and higher indexed Riemann sheets. These poles are associated with
    resonances in the $K\pi$, $K\pi\pi$, $K\eta$ production channels. We summarize these resonances and their properties in \cref{tab:hmes:resonances}.

    \item \label{prop:hmes:properties:large-momentum-limit}
    In the limit $|q^2| \to \infty$, the form factor $f_+(q^2) \sim 1/q^2 \to 0$, on account of the exchange of hard gluon interactions
    dominating the transition~\cite{Lepage:1979zb}. The same argument implies $f_0(q^2) \sim 1/q^2 \to 0$ as $q^2\to \infty$.
\end{properties}
A visual summary of the relevant phase space and pole properties is depicted in \cref{fig:hmes:illustration}.

\subsection{Dispersive Bounds}
\label{sec:hmes:bounds}

Dispersive bounds for hadronic form factors, developed first in the context of semileptonic
kaon decay~\cite{Ling-Fong:1971tjw,Okubo:1971my}, have now been firmly established for more than half
a century in the context of heavy-to-heavy meson form factors~\cite{Boyd:1995cf,Caprini:1997mu}.
The bounds on a form factor $F$ take the schematic form
\begin{equation}
    \label{eq:hmes:basic-bound}
    \int_{q^2 = \tp}^{\infty} \omega_F (q^2) |F|^2 d q^2 \leq 1
\end{equation}
where $\omega$ is a positive-definite weight function determined in the derivation of the bound.
It is closely related to the outer functions to be defined below.

In the modern formulation, the standard parameterisation of dispersively-bounded form factors
for evaluation in the semi-leptonic region is
\begin{equation}
    \label{eq:hmes:bounds:old-param}
    \begin{aligned}
        f_+(q^2)
            & = \frac{1}{\phi_+(q^2)} \sum_{k=0}^{N_+} a^{(+)}_k \left[z(q^2; t_0)\right]^k\,,
            &
        f_0(q^2)
            & = \frac{1}{\phi_0(q^2)} \sum_{k=0}^{N_0} a^{(0)}_k \left[z(q^2; t_0)\right]^k\,.
    \end{aligned}
\end{equation}
In the above expressions, we have made use of the conformal mapping
\begin{equation}
    q^2 \mapsto z(q^2; t_s) = \frac{\sqrt{\tp - q^2} - \sqrt{t_s - q^2}}{\sqrt{\tp - q^2} + \sqrt{t_s - q^2}}\,,
\end{equation}
with $t_s$ an arbitrary parameter that fulfils $t_s < \tp$.
The parameter $t_0$ used in \cref{eq:hmes:bounds:old-param} can be chosen freely.
Here and throughout this work, we choose $t_0 = \qty{-4}{\GeV^2}$.
The above change of variable maps the form factors' first Riemann sheet onto the unit open disk, their second
Riemann sheet onto the complement of the unit disk, and the branch cut connecting both
sheets onto the unit circle.
The expansion coefficients $a^{(+)}_k$ and $a^{(0)}_k$ are not independent by virtue
of the identity $f_+(0) = f_0(0)$.
The prefactors in \cref{eq:hmes:bounds:old-param} involve the outer functions $\phi_+$ and $\phi_0$.
For $\bar{u} s$ currents these are given by
\begin{equation}
    \begin{aligned}
        \phi_+ (t)
            & = \left| \frac{dz}{dt} \right|^{-1/2} \sqrt{\frac{1}{32 \pi \chi_{1^-} (-Q^2)}} \left[\frac{t - \tp}{-z(t; \tp)} \frac{t-\tm}{-z(t;\tm)}\right]^{3/4} \times \frac{-z(t;0)}{t} \left(\frac{-z(t; -Q^2)}{t+Q^2}\right)^{3/2}
            \\
        \phi_0 (t)
            & = \left| \frac{dz}{dt} \right|^{-1/2} \sqrt{\frac{3}{32 \pi \chi_{0^+} (-Q^2)}} \sqrt{\tp \tm} \left[\frac{t - \tp}{-z(t; \tp)} \frac{t-\tm}{-z(t;\tm)}\right]^{1/4} \times \frac{-z(t;0)}{t} \frac{-z(t; -Q^2)}{t+Q^2} \,,
    \end{aligned}
\end{equation}
within which the $\chi_{1^-}$ and $\chi_{0^+}$ are normalisations. The normalisations are obtained from the local operator product expansion
of a two-point correlation function with momentum transfer $-Q^2 < 0$.
In this work, we use $Q^2 = \qty{4}{\GeV^2}$, evaluating the expressions provided in Ref.~\cite{Hill:2006bq} with $\alpha_s (\qty{2}{\GeV}) = 0.307$, $m_s = \qty{93.5 \pm 0.8}{\MeV}$~\cite{ParticleDataGroup:2024cfk}, and condensate values from Ref.~\cite{Bordone:2019guc}. We obtain
\begin{equation}
    \label{eq:hmes:susceptibility}
    \begin{aligned}
        \chi_{1^-} (Q^2 & = \qty{4}{\GeV^2}) = \qty{3.446 \pm 0.009 e-3}{\GeV^{-2}}
        \\
        \chi_{0^+} (Q^2 & = \qty{4}{\GeV^2}) = \num{2.07 \pm 0.41 e-4} \,.
    \end{aligned}
\end{equation}
The perturbative convergence of $\chi_{0^+}$ is poor, and we have therefore assigned an uncertainty of \qty{20}{\percent},
larger than what one obtains as a nominal parametric uncertainty.

The dispersive bound on the form factors as given in \cref{eq:hmes:basic-bound} then implies a simple bound on the expansion parameters $a_k$
\begin{equation}
\label{eq:hmes:simple-bound}
    \begin{aligned}
        \sum_{k=0}^{\infty} \left|a^{(+)}_k\right|^2
            & \leq 1\,,
            &
        \sum_{k=0}^{\infty} \left|a^{(0)}_k\right|^2
            & \leq 1\,.
    \end{aligned}
\end{equation}

\subsection{Improved Parametrisation}
\label{sec:hmes:param}

The dispersively-bounded parametrisation for the $K\to \pi$ form factors given in \cref{eq:hmes:bounds:old-param}
respects \cref{prop:hmes:properties:no-bound-states,prop:hmes:properties:real-below-threshold}.
However, it does not satisfy the remaining properties.
The specific issues are:
\begin{enumerate}
    \item \Cref{prop:hmes:properties:finite-threshold,prop:hmes:properties:angular-momentum-threshold} are violated, since the outer functions leads to a divergence in both $f_+$ and $f_0$ at threshold, while the imaginary part of $f_+$ develops as $(\tp - q^2)^{3/2}$ above threshold.
    The effect of violating this property has been discussed in the context of the pion form factor~\cite{Buck:1998kp},
    and of the $B\to \pi$ vector form factor~\cite{Becher:2005bg}.
    In the context of the latter, it inspired a new parametrisation thereof~\cite{Bourrely:2008za}.
    The concerns of Refs.~\cite{Buck:1998kp,Becher:2005bg,Bourrely:2008za} apply here as well:
    since the endpoint of the \Kl3 decay $t_- \simeq \qty{0.13}{\GeV^2}$ and the
    branch point $t_+ \simeq \qty{0.40}{\GeV^2}$ are very close to each other, mismodelling of the
    form factor in the vicinity of the branch point can have undue effect on the shape of the form factors in the \Kl3 phase space
    and/or hinder the description of the \Ptau decay.

    \item \Cref{prop:hmes:properties:further-cuts} is violated, since the mapping from $q^2$ to $z$ introduces
    only a single, square-root-like branch cut.
    We do not attempt to improve on this issue here and refer to Ref.~\cite{Balz:2025auk} for a prospective solution.

    \item \Cref{prop:hmes:properties:above-threshold-resonances} is violated, in so far that a pattern of multiple resonances cannot be described by the truncated series expansion in $z$.
    Even for a single resonance, this has not been shown to be possible;
    see the discussion in Ref.~\cite{Buck:1998kp} where on the order of 60 parameters
    are required to approximately fit the $\rho$ peak while producing unphysical features near the $\pi\pi$ threshold.

    \item \Cref{prop:hmes:properties:large-momentum-limit} is violated, since one finds for the asymptotic behaviour $f_+ (q^2) \sim (q^2)^{1/4}$ and $f_0(q^2) \sim (q^2)^{3/4}$ as $q^2 \to \infty$.
\end{enumerate}
Three of these issues (\#1, \#3, and \#4) have been previously discussed and addressed by means of a new
and simple parametrisation in the context of the pion form factor~\cite{Kirk:2024oyl}.
Here, we propose to apply those modifications to the $K\to \pi$ form factors, generalising that work to account for the existence of multiple above threshold poles.
To summarise the changes to the standard parametrisation made in that work, we:
\begin{itemize}
    \item include an overall weight function that removes the nominal pole at threshold, and generates the correct asymptotic behaviour for $q^2 \to \infty$, to ensure \cref{prop:hmes:properties:finite-threshold,prop:hmes:properties:large-momentum-limit}.
    The weight functions are simplest to describe in $z$ space, taking the forms $W_+(z) = (1+z)^{2} (1-z)^{5/2}$ and $W_0(z) = (1+z)^{1} (1-z)^{7/2}$.    
    \item explicitly account for above-threshold resonances, which are located outside the unit circle (corresponding to the second Riemann sheet as required by \cref{prop:hmes:properties:above-threshold-resonances}) at
    $$
        z_r = z( (M_r - i \Gamma_r / 2)^2; t_0)^{-1}
    $$
    for a resonance with mass $M_r$ and width $\Gamma_r$.
    Each resonance is included as a conjugate pair through a factor $1 / (z - z_r)(z - z_r^*)$, so as to maintain \cref{prop:hmes:properties:real-below-threshold}.
\end{itemize}
The new parameterisation then reads
\begin{equation}
    \label{eq:hmes:bounds:param}
    \begin{aligned}
        f_+(q^2)
            & = \frac{W_+(q^2)}{\phi_+(q^2)} \prod_{r=1}^{M_+} \frac{1}{z - z_{r,+}} \frac{1}{z - z_{r,+}^*} \sum_{k=0}^{N_+} \bp{k} \left[z(q^2; t_0)\right]^k\,,
        \\
        f_0(q^2)
            & = \frac{W_0(q^2)}{\phi_0(q^2)} \prod_{r=1}^{M_0} \frac{1}{z - z_{r,0}} \frac{1}{z - z_{r,0}^*} \sum_{k=0}^{N_0} \bz{k} \left[z(q^2; t_0)\right]^k\,.
    \end{aligned}
\end{equation}
Two linear relations exist between the expansion coefficients introduced in \cref{eq:hmes:bounds:param}.

The first relation reads
\begin{equation}
    \label{eq:hmes:bounds:relation-1}
    \bp{0} = -\frac{1}{X'} \left( \sum_{n=1}^{N_+} (-1)^n \bp{n} (X' - n X) \right)\,.
\end{equation}
where we use
\begin{equation}
    \begin{aligned}
        X(z)
            & = \frac{W_+(z)}{\phi_+(z)} \prod_{r=1}^{M_+} \frac{1}{z - z_{r,+}} \frac{1}{z - z_{r,+}^*}\,,
            &
        X'
            & = \frac{dX}{dz}\,.
    \end{aligned}
\end{equation}
By construction, the contributions to the imaginary part arise from odd powers in $(z + 1) \sim (q^2 - \tp)^{1/2}$.
\Cref{eq:hmes:bounds:relation-1} ensures that the leading contribution to the imaginary part of $f_+$, which behaves
as $\propto (q^2 - \tp)^{1/2}$, vanishes.
As a consequence, $\Im f_+(q^2 = \tp) \propto (q^2 - \tp)^{3/2}$ as required
by \cref{prop:hmes:properties:angular-momentum-threshold}.

The second relation reads
\begin{equation}
    \bz{0}
        = \frac{\tilde{\phi}_0 (z_0)}{\tilde{\phi}_+ (z_0)} \frac{\Pi_{r=1}^{M_0}|z_0 - z_{r,0}|^2}{\Pi_{r=1}^{M_+} |z_0 - z_{r,+}|^2} \sum_{n=0}^N \bp{n} z_0^n -  \sum_{n=1}^N \bz{n} z_0^n\,,
\end{equation}
where we abbreviate $z_0 = z(0;t_0)$. This relation ensures that $f_+(0) = f_0(0)$.\\

With this parameterisation, we find ourselves unable to express the dispersive bound in a closed form,
\ie, similar to \cref{eq:hmes:simple-bound} involving the expansion coefficients.
However, we can still determine the saturation of the dispersive bound due to the $K\pi$ form factors by numerically evaluating the integral
\begin{equation}
\label{eq:hmes:numerical_bound}
    \textrm{saturation}_{+,0} \equiv \frac{1}{2\pi} \int_{-\pi}^{+\pi} d \vartheta \left|\phi_{+,0}(q^2(z)) f_{+,0}(q^2(z))\right|^2\bigg|_{z = e^{i\vartheta}}\,.
\end{equation}
The quantities $\chi_{1^-}$ and $\chi_{0^+}$ (and therefore the normalisation of the bound saturations)
feature sizeable uncertainties as indicated in \cref{eq:hmes:susceptibility}.

\section{Analysis Setup}
\label{sec:setup}

We perform a Bayesian analysis of the available data on \Kl2, \Kl3, $\tau \to K\pi\nu$ and $\tau \to K \nu$ decays.
The central quantity for this analysis is the so-called posterior probability density function (PDF):
\begin{equation}
    \label{eq:setup:posterior}
    P(\vecx\,|\, D, M) \propto P(D \,|\, \vecx, M)\, P_0(\vecx \,|\, M)\,.    
\end{equation}
In the course of our analysis, we either optimize this posterior PDF with respect to the parameters $\vecx$
or we draw random samples $\vecx \sim P(\vecx\,|\, D,M)$.
The former provides for a best-fit point while the latter provides
insight into the extend of the allowed parameter space and is crucial input to producing posterior predictions.
The set of varied parameters $\vecx$ is tied to the underlying fit model $M$.
The fit models used in this analysis are discussed in \cref{sec:setup:priors}.
The posterior PDF (\cref{eq:setup:posterior}) is expressed in terms of the likelihood function
$P(D \,|\, \vecx, M)$ and the prior PDF $P_0(\vecx \,|\, M)$.
The former accounts for the experimental and theoretical constraints following from any of the labelled data sets $D$,
which are discussed in \cref{sec:setup:data}.
The latter accounts for our a-priori knowledge of the model parameters; see
\cref{sec:setup:priors} for a discussion of the models, their parameters, and their prior PDFs.
\\

Our analysis is carried out by using the \EOS software~\cite{EOSAuthors:2021xpv} in version 1.0.19~\cite{EOS:v1.0.19}.
To draw the posterior samples, \EOS uses dynamical nested sampling~\cite{Higson:2018} as implemented in the
open-source \dynesty software~\cite{dynesty:v2.0.3}. The use of dynamical nested sampling ensures simultaneously
high accuracy for the posterior predictions and the estimate of the evidence.
The contents of the remainder of this section provide a detailed explanation of the analysis file used in the course
of this study. The analysis file is available as part of the supplementary material~\cite{EOS-DATA-2025-05}.

\subsection{Experimental Data}
\label{sec:setup:data}

Our analysis uses the following experimental results for exclusive processes mediated by the $us\ell\nu$ sector of the WET:
\begin{description}
    \item[$\boldsymbol{K^-\to \mu^- \bar\nu}$]
    We use the PDG world average~\cite{ParticleDataGroup:2024cfk}
    of branching ratio measurements of the muon mode, which reads
    \begin{equation}
        \mathcal{B}(K^-\to \mu^-\bar\nu)_\text{PDG 2024} = (63.60 \pm 0.16)\%\,.
    \end{equation}
    This average includes measurements by the KLOE and OSPK experiments~\cite{KLOE:2005xes,Chiang:1972rp}
    and provides a total of one observation to our analysis.

    The corresponding electron mode has, to the best of our knowledge, only been measured by means of the
    ratio $\Gamma(K^-\to e^-\bar\nu)/\Gamma(K^-\to \mu^-\bar\nu)$.
    Including it in the fit would therefore induce (potentially large) correlations between the measurement of this ratio and
    the muon mode measurement. Hence, we do not include the electron mode measurement in our analysis.

    \item[$\boldsymbol{K^-\to \pi^0 \lbrace e^-, \mu^-\rbrace \bar\nu}$]
    Two sets of measurements of the branching ratios of these decays enter the world average~\cite{ParticleDataGroup:2024cfk}:
    a joint measurement of both decay modes by KLOE~\cite{KLOE:2007wlh}, and a measurement of each mode without known
    correlations~\cite{Chiang:1972rp}. The average is clearly dominated by the KLOE measurement
    \begin{equation}
    \begin{aligned}
        \mathcal{B}(K^- \to \pi^0 e^- \bar{v})_\text{KLOE}
            & = \qty{4.965 \pm 0.038 \pm 0.037}{\percent} \\
        \mathcal{B}(K^- \to \pi^0 \mu^- \bar{v})_\text{KLOE}
            & = \qty{3.233 \pm 0.029 \pm 0.026}{\percent}
    \end{aligned}
    \end{equation}
    which features a linear correlation coefficient of $\rho = 0.627$.
    To avoid issues with ignoring the unknown correlations for the measurement presented in Ref.~\cite{Chiang:1972rp},
    we use only the KLOE measurement, which contributes two observations to our analysis.

    \item[$\boldsymbol{\KS \to \pi^+\lbrace e^-, \mu^-\rbrace \bar\nu}$] 
    The world average includes a measurement of the branching ratio for $K_S \to \pi^+ e^- \bar\nu$ by KLOE~\cite{KLOE:2002lao}
    \begin{equation}
        \mathcal{B}(\KS \to \pi^+ e^- \bar\nu)_\text{KLOE} = \qty{3.455 \pm 0.17 \pm 0.075 e-2}{\percent}
    \end{equation}
    and a determination of the same branching ratio extracted from the ratio
    $\Gamma(\KS \to \pi^+ e^-\bar\nu) / \Gamma(\KL \to \pi^+ e^-\bar\nu)$.
    To the accuracy of our analysis, the latter ratio is fully determined.
    Hence, we use only the KLOE measurement~\cite{KLOE:2002lao},
    which contributes one observation to our analysis.\\

    The world average does not yet include a measurement of the branching ratio
    for the muon mode $K_S \to \pi^+ \mu^- \bar\nu$ by the
    KLOE-2 experiment~\cite{KLOE-2:2019rev}
    \begin{equation}
        \mathcal{B}(\KS \to \pi^+ \mu^- \bar\nu)_\text{KLOE-2} = \qty{2.28 \pm 0.055 \pm 0.085 e-2}{\percent}
    \end{equation}
    We use this measurement, which contributes one observation to our analysis.

    \item[$\boldsymbol{\KL \to \pi^+\lbrace e^-, \mu^-\rbrace \bar\nu}$]
    The world average~\cite{ParticleDataGroup:2024cfk} is determined from two individual correlated measurements
    of the branching ratios for $K_L \to \pi^+ \ell \bar\nu$, with $\ell = e,\mu$.
    The measurement by the KLOE experiment~\cite{KLOE:2005vdt} reads
    \begin{equation}
        \begin{aligned}
            \mathcal{B}(\KL \to \pi^+ e^- \bar{v})_\text{KLOE}
                & = \qty{20.035 \pm 0.025 \pm 0.075}{\percent} \\
            \mathcal{B}(\KL \to \pi^+ \mu^- \bar{v})_\text{KLOE}
                & = \qty{13.49 \pm 0.025 \pm 0.075}{\percent}
        \end{aligned}
    \end{equation}
    with a correlation coefficient $\rho = -0.25$.
    When using the above, we combine the given statistical and systematic uncertainties in quadrature.
    The measurement by the KTeV experiment~\cite{KTeV:2004hpx} reads
    \begin{equation}
        \begin{aligned}
            \mathcal{B}(\KL \to \pi^+ e^- \bar{v})_\text{KTeV}
                & = \qty{20.335 \pm 0.055}{\percent} \\
            \mathcal{B}(\KL \to \pi^+ \mu^- \bar{v})_\text{KTeV}
                & = \qty{13.505 \pm 0.045}{\percent}
        \end{aligned}
    \end{equation}
    with a correlation coefficient $\rho = +0.15$.
    These two joint determinations of the branching ratios are mutually compatible only at the \qty{2.7}{\sigma} level, with the incompatibility primarily in the electron mode.
    In the reported PDG world averages, the electron mode and the muon mode are combined individually,
    applying the standard PDG averaging procedure to each mode.
    This requires inflating the reported error on the electron average by a scale factor $S = 3.1$,    
    while the error on the muon average is unchanged~\cite{ParticleDataGroup:2024cfk}.
    While there is no widely agreed on procedure for combining discrepant multivariate measurements, we choose to apply a PDG-like scale factor to the joint measurements from both experiments, inflating both electron and muon uncertainties by the common scale factor $S = 2.1$.

    \item[$\tau^- \to K^- \nu$] We use the PDG world average~\cite{ParticleDataGroup:2024cfk}
    for the branching ratio
    \begin{equation}
        \mathcal{B}(\tau^- \to K^- \nu)_\text{PDG 2024} = \qty{0.685 \pm 0.023}{\percent}\,,
    \end{equation}
    which includes measurements from the ALEPH~\cite{ALEPH:1999jxs}, CLEO~\cite{CLEO:1994bvq},
    DELPHI~\cite{DELPHI:1994gzt}, and OPAL~\cite{OPAL:2000fde} experiments.
    It contributes a total of one observation to our analysis.

    \item[$\boldsymbol{\tau^- \to \KS \pi^- \nu}$] We use the PDG world average~\cite{ParticleDataGroup:2024cfk}
    for the branching ratio
    \begin{equation}
        \mathcal{B}(\tau^- \to \KS \pi^- \nu)_\text{PDG 2024} = \qty{0.4195 \pm 0.011}{\percent}\,,
    \end{equation}
    which includes measurements by the ALPEH~\cite{ALEPH:1999jxs}, Belle~\cite{Belle:2014mfl}, L3~\cite{L3:1995cos},
    and OPAL~\cite{OPAL:1999bbs} experiments. This average includes a scale factor of 1.5 and contributes
    one observation to our analysis.

    In addition, we include a measurement of the normalized decay rate
    \begin{equation}
        P(\tau^-\to \KS \pi^- \nu) \equiv \frac{1}{\Gamma(\tau^-\to \KS \pi^- \nu)} \frac{d\Gamma(\tau^-\to \KS \pi^- \nu)}{dq^2}
    \end{equation}
    by the Belle experiment~\cite{Belle:2007goc} (this data can be seen in \cref{fig:res:tau-decay}). It contributes 100 observations to our analysis.

    \item[$\boldsymbol{\tau^- \to K^- \pi^0 \nu}$] We use the PDG world average~\cite{ParticleDataGroup:2024cfk}
    for the branching ratio
    \begin{equation}
        \mathcal{B}(\tau^- \to K^- \pi^0 \nu) = \qty{0.426 \pm 0.016}{\percent}\,,
    \end{equation}
    which includes measurements by the ALEPH~\cite{ALEPH:1999jxs}, BaBar~\cite{BaBar:2007yir},
    CLEO~\cite{CLQCD:2023sdb}, and OPAL~\cite{OPAL:2004icu} experiments.
    This average contributes one observation to our analysis.
\end{description}

In the following, we will refer to combinations of the above as these named data sets. These are defined as:
\begin{description}
    \item[\texttt{3body}] All 3-body decay modes, \ie, the branching ratios for all $K \to \pi \ell \nu$ decays, as well as the branching ratios of both $\tau \to K \pi \nu$ modes and the differential measurement from Belle.
    It accounts for a total of 110 observations.
    \item[\texttt{all}] The combination of the datasets \texttt{3body} with the branching ratios for the $\tau \to K \nu$ and $K \to \mu \nu$ decays.
    It accounts for a total of 112 observations.
\end{description}

\subsection{Fit models, Parameters of Interest, Nuisance Parameters, and their Priors}
\label{sec:setup:priors}

We use the following fit models:
\begin{description}
    \item[FF] In this model, the parameters of interest are those of our new $K \to \pi$ form factors given in \cref{eq:hmes:bounds:param}. Here we fix $\absV{us} = 0.22515$ which is the value obtained using $\Kmutwo/ \pi_{\mu 2}$ + unitarity~\cite{ParticleDataGroup:2024cfk,PDG:Vud_Vus_review}.
    This fit model is exclusively used in conjunction with the \texttt{3body} experimental dataset.
    For imposing the Callan-Treiman relations, we vary the parameters $f_{K}$, $f_{\pi}$, $\Delta_{\textrm{CT}}$, and $\Delta_{\widetilde{\textrm{CT}}}$ within informative gaussian priors, which are discussed below.
    For the description of the $K\pi$ form factors, it features additionally $2 (M_+ + M_0) + (N_+ + N_0 - 2)$ real-valued parameters. Unless otherwise stated, we choose uniform priors for each of these
    parameters. The intervals are chosen wide enough to avoid cutting into the peak due to the likelihood. For details on the
    prior ranges and choices of the model meta parameters $M_+$, $M_0$, and $N_+ = N_0$,
    we refer to the analysis file \texttt{analysis\_FF.yaml} that is part of the supplementary
    material~\cite{EOS-DATA-2025-05}.
    \item[CKM] This model is based on the FF model, except that we additionally float $|\V{us}|$
    in the range $[0.223, 0.227]$ as the new parameter of interest and utilise the
    lattice QCD result for the semileptonic kaon form factors at $q^2 = 0$ as an additional theoretical input.
    This fit model is exclusively used with the dataset \texttt{all}.
    For further details, we refer to the analysis file \texttt{analysis\_CKM.yaml} in the supplementary material.
    \item[BSM] This model is based on the FF model. We use it to study the potential size of BSM effects
    in the data by introducing two new degrees of freedom
    in the form of the WET Wilson coefficients \CVL and \CVR. For simplicity we assume both to be real-valued quantities.
    that are varied in the ranges
    \begin{equation}
        \begin{aligned}
                & 1.000 \leq \CVL \leq 1.025\,,
                &
                & -0.01\leq \CVR \leq 0.02\,.
        \end{aligned}
    \end{equation}
    The new parameters are varied alongside the same hadronic FF parameters as in the FF model.
    Further hadronic matrix elements only arise if BSM operators with $\bar{u} \sigma^{\mu\nu} s$ currents are included in the analysis.
    Since we only consider vector type currents in this fit model, we are able
    to consistently fit the form factor from data while simultaneously quantifying potential BSM effects.
    To provide an absolute scale for the magnitude of these Wilson coefficients, we fix $\absV{us} = 0.22515$, as in the FF fit model.
    We assume lepton-flavour universality of the Wilson coefficients.
    As in the CKM model, we additionally constrain the semileptonic kaon form factors based on the lattice QCD world average.
    For further details, we refer to the analysis file \texttt{analysis\_BSM.yaml} in the supplementary material.
    This fit model is exclusively used with the dataset \texttt{all}.
\end{description}

The informative priors referred to in the description of the FF model are chosen as follows:
\begin{description}
    \item[$\boldsymbol{K^-\to \mu^-\bar\nu}$] The hadronic input to leptonic decays of a pseudoscalar meson
    is expressed in the SM by means of a single scalar parameter, the kaon decay constant
    \begin{equation}
        f_{K} = \qty{155.7(0.3)}{\MeV}\,.
    \end{equation}
    from the 2024 FLAG report~\cite{FlavourLatticeAveragingGroupFLAG:2024oxs}, based on lattice
    QCD results provided in Refs.~\cite{ExtendedTwistedMass:2021qui,Carrasco:2014poa,Bazavov:2014wgs,Dowdall:2013rya}.
    \item[$\bar{K}^0\to \pi^+\ell^-\nu$] The hadronic input to the semileptonic decays of a kaon includes the
    normalisation of the $f_+$ form factor at $q^2 = 0$. The 2024 FLAG report~\cite{FlavourLatticeAveragingGroupFLAG:2024oxs}, based on results by the ETMC~\cite{Carrasco:2016kpy} and FNAL/MILC~\cite{FermilabLattice:2018zqv} collaborations provides the average
    \begin{equation}
    \label{eq:setup:priors:fpzero}
        f_+(0) = 0.9698 \pm 0.0017\,.
    \end{equation}
    \item[Callan-Treiman] The Callan-Treiman theorem depends on the ratio of the kaon and pion decay constants.
    In the latest FLAG report~\cite{FlavourLatticeAveragingGroupFLAG:2024oxs} this ratio is given to \qty{0.16}{\percent} precision, while the pion decay constant is only known to \qty{0.6}{\percent} (and as given above, the kaon decay constant is known to \qty{0.19}{\percent}).
    We therefore choose the more conservative option by treating the pion and kaon decay constants as separate nuisance parameters, using the above kaon value and 
    \begin{equation}
        f_{\pi} = \qty{130.2 (0.8)}{\MeV} 
    \end{equation}
    for the pion decay constant likelihood (based on lattice results provided in Refs.~\cite{Follana:2007uv,MILC:2010hzw,RBC:2014ntl,CLQCD:2023sdb}).
    The corrections $\Delta_\text{CT}$ and $\Delta_{\widetilde{\text{CT}}}$ to the Callan-Treiman relations
    are included via two uncorrelated Gaussian likelihoods using the numerics in \cref{eq:DeltaCT-params}.
\end{description}
For all fit models we impose the Callan-Treiman relations and the dispersive bounds.
The latter are imposed by numerically computing the saturation and penalizing parameter points for which
the saturation exceeds $\qty{100}{\percent}$.
To so, we use eq. (A3) of Ref.~\cite{Bordone:2019vic} to impose a smooth penalty function that
accounts for the relative uncertainties of $\chi_{1^-}$ and $\chi_{0^+}$.

\section{Results}
\label{sec:results}

\subsection{Fits Extracting the Hadronic Matrix Elements}
\label{sec:results:HME}

We have conducted a variety of fits using the FF model, either fixing individual resonance parameters according to their PDG values or letting them float.
We successfully fit the available data \texttt{3body}, which excludes any theoretical information on the $K\to \pi$
form factors in the semileptonic region. We accept fits that feature a $p$ value in excess of
$\qty{3}{\%}$, which is our a-priori threshold. Amongst these accepted fits,
the worst performing is $M_+ = M_0 = 1$ and $N_+ = N_0 = 3$ at $p \simeq \qty{6}{\%}$,
\ie, accounting for the $\Kresonance{892}$ and the $\Kzeroresonance{700}$ only.
With respect to the resonances appearing in the \Ptau decay, our finding can be summarised as follows:
\begin{enumerate}
    \item We are able to fit the resonance parameters (mass and width) of the \Kresonance{892} from the available data.
    To do this, we fix the parameters relating to the remaining resonances in the $f_+$ and $f_0$ form factors.
    We obtain pole parameters for the \Kresonance{892},
    \begin{equation}
    \begin{aligned}
        M_{892}
            & = \qty[uncertainty-descriptors = {stat, model}]{891.87 \pm 0.18 \pm 0.05}{\MeV} \,,
            \\
        \Gamma_{892}
            & = \qty[uncertainty-descriptors = {stat, model}]{45.5 \pm 0.4 \pm 0.5}{\MeV} \,,
    \end{aligned}
    \end{equation}
    that are in a good agreement with the PDG values. We discuss how we assess the model uncertainty below.
    
    \item We are unable to fit the resonance parameters of the \Kzeroresonance{700} from the available data, regardless
    of our choice of treatment of the remaining resonances and their parameters.

    \item Fixing the \Kzeroresonance{700} resonance parameters to those of the PDG, we are able to fit \emph{either}
    the \Kresonance{1410} \emph{or} the \Kzeroresonance{1430} resonance parameters, but not both at the same time.
    Given the closeness of these two resonances, we speculate that separating them from each other is not possible
    given the available data.
    Since they differ in their total angular moment, we hold that additional information in form of the angular observable
    arising from the $K\pi$ helicity angle has the potential to change this outcome, \emph{i.e.}, to cleanly
    separate the two resonances from each other and fit their parameter simultaneously.
    
    \item We are unable to fit the resonance parameters of the \Kresonance{1680} from the available data.
    This is likely related to the fact that it is situated very close to the endpoint of the \Ptau decay phase space
    $\sqrt{q^2} \leq m_\tau \simeq \qty{1.777}{\GeV}$, which causes a small decay rate in the region close to \Kresonance{1680}.
\end{enumerate}

\begin{figure}
    \includegraphics[width=0.9\linewidth]{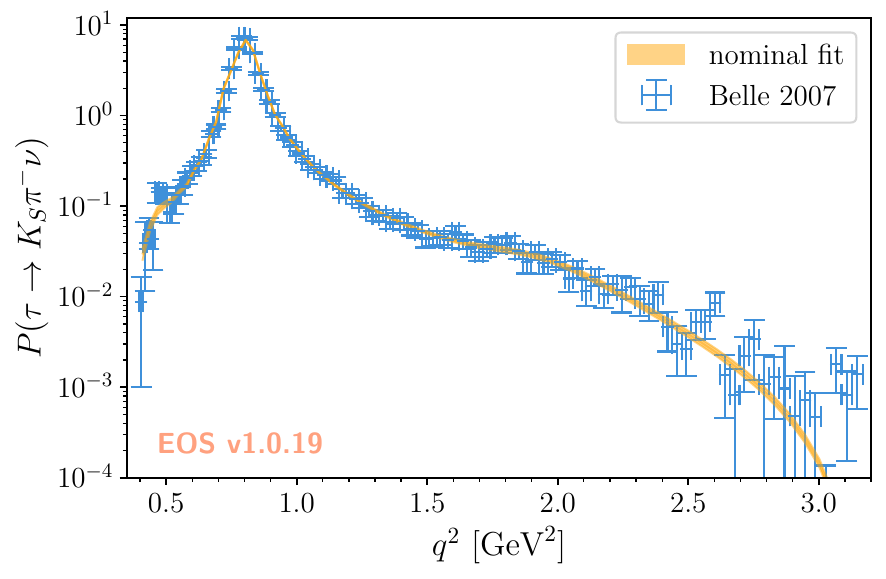}
    \caption{Normalised differential \Ptau decay data from Belle~\cite{Belle:2007goc} (blue error bars), along with the posterior prediction and \qty{68}{\percent} probability envelope of our nominal model fit (orange band).}
    \label{fig:res:tau-decay}
\end{figure}

In light of our findings, we choose to use as our nominal result a model containing \Kresonance{892}, \Kresonance{1410}, \Kzeroresonance{700} and \Kzeroresonance{1430} (i.e.\ $M_+ = M_0 = 2$ in \cref{eq:hmes:bounds:param}), as such a model contains the four resonances we know to be present and to which the available data is sensitive.
We find that going to at least order $z^3$ in the expansion is needed to produce an acceptable p-value, and therefore we go one order higher to ensure the systematic truncation uncertainties are under control (i.e.\ we use $N_+ = N_0 = 4$).
At higher orders in the expansion, we see degeneracies appearing, indicating that the parameters are poorly constrained.
This model fit has a p-value of \qty{72}{\percent}, and the posterior prediction for the differential \Ptau decay rate is shown in \cref{fig:res:tau-decay}.
We additionally use a model that also contains the \Kresonance{1680} (i.e.\ $M_+ = 3$, $M_0 = 2$).
Here we find that going to $z^4$ in the expansion is required for an acceptable p-value, and again use the next higher order to study truncation errors.
The difference between (pseudo-)observables in these four representative models
\begin{equation}
    \label{eq:results:hmes:metaparameters}
    \lbrace (M_+, M_0, N_+, N_0) : (2,2,3,3), (2,2,4,4), (3,2,4,4), (3,2,5,5) \rbrace
\end{equation}
is used as our estimate of the model uncertainty.

\subsubsection{\texorpdfstring{$K\to \pi$}{K to pi} Form Factor}
Our fits produce posterior-predictive results for the $K\to \pi$ form factor.
We first compare the value at zero momentum transfer $q^2 = 0$.
From our representative fit models as given in \cref{eq:results:hmes:metaparameters}, we obtain
\begin{equation}
    f_+(0) = \num[uncertainty-descriptors = {stat, model}]{0.9713 \pm 0.0019 \pm 0.0030}
\end{equation}
This result should be compared to the world average of the $K\to \pi$ form factor from lattice QCD determinations~\cite{FlavourLatticeAveragingGroupFLAG:2024oxs},
which is provided in \cref{eq:setup:priors:fpzero}.
We find excellent agreement between our result and the world average of lattice QCD predictions.

Experimental studies of the $\Kl3$ decay have access to the shape of the decay
distribution and therefore access to the shape of the form factors.
This shape is probed through a model that Taylor expands the form factors about $q^2 = 0$,
\begin{equation}
    f_{+,0} (t) = f_{+,0} (0) \left( 1 + \frac{\lambda^\prime_{+,0} t}{m_\pi^2} + \frac{\lambda^{\prime\prime}_{+,0} t^2}{2 m_\pi^4} \right) \,.
\end{equation}
Our model is not limited to the first two terms in the expansion. However, restricting
the degrees of freedom in the Taylor expansion above is liable to introduce bias in the results\footnote{%
    As evidenced by the substantial differences in the results for $\lambda_+^\prime$ from a linear and a quadratic
    fit in the PDG world average~\cite{ParticleDataGroup:2024cfk}.
}.
Here, we compare the results for the linear coefficient from a joined analysis of the electron
and muon mode $\lambda_+^\prime$~\cite{NA482:2018rgv}
\begin{equation}
    \begin{aligned}
        \lambda^\prime_+\big|_\text{NA48/2}^\text{quadratic model}
            & = \num[uncertainty-descriptors = {stat, sys}]{0.02424 \pm 0.00075 \pm 0.00130}\,,
            \\
        \lambda^\prime_0\big|_\text{NA48/2}^\text{linear model}
            & = \num[uncertainty-descriptors = {stat, sys}]{0.01447 \pm 0.00063 \pm 0.00117}\,,
    \end{aligned}
\end{equation}
with our posterior prediction
\begin{equation}
    \begin{aligned}
        \lambda^\prime_+
            & = \num[uncertainty-descriptors = {stat, model}]{0.0238 \pm 0.0007 \pm 0.0020}\,,
            \\
        \lambda^\prime_0
            & = \num[uncertainty-descriptors = {stat, model}]{0.0119 \pm 0.0011 \pm 0.0010}\,.
    \end{aligned}
\end{equation}
We find excellent agreement at the levels of \qty{0.17}{\sigma} and \qty{1.29}{\sigma}, respectively.

\subsubsection{Predictions for $D$ Decays}

Another use case for these form factor results comes from hadronic $D$ decay studies, where the topological amplitudes
of the annihilation type can (up to corrections of order $1/N_c^2$) be written in terms of the $D$ decay constant and the scalar $K \to \pi$ form factor evaluated at $q^2 = m_D^2$ or $q^2 = m_{D_s}^2$~\cite{Muller:2015lua}.

In Ref.~\cite{Muller:2015lua}, the authors assume $f_0(m_{D_s}^2) = f_0(m_D^2)$, and take an estimate based on extrapolations
in Ref.~\cite{Belle:2007goc}, which reads
\begin{equation}
1 \lesssim | f_0(m_{D_{(s)}}^2)| \lesssim 4.5 \,, \quad -\pi \lesssim \arg f_0(m_{D_{(s)}}^2) \lesssim \pi
\end{equation}
as a suitable range to describe the available \Ptau decay data.
Using our results, we are able to predict the absolute value more precisely
\begin{equation}
\label{eq:results:Ddecay_f0}
|f_0(m_D^2)| \approx |f_0(m_{D_s}^2)| = \num[uncertainty-descriptors = {stat, model}]{1.7 +- 0.7 +- 0.2} \,,
\end{equation}
while for the phase our fit still provides no sensitivity. Both quantities are strongly correlated, with
the correlation coefficient close to unity. For the ratio between both quantities we find
\begin{equation}
    |f_0(m_D^2)| / |f_0(m_{D_s}^2)| \simeq 1 \pm 0.2\,.
\end{equation}

\subsection{Fits Extracting $\absV{us}$}
\label{sec:results:Vus}

\begin{figure}
    \includegraphics[width=0.8\linewidth]{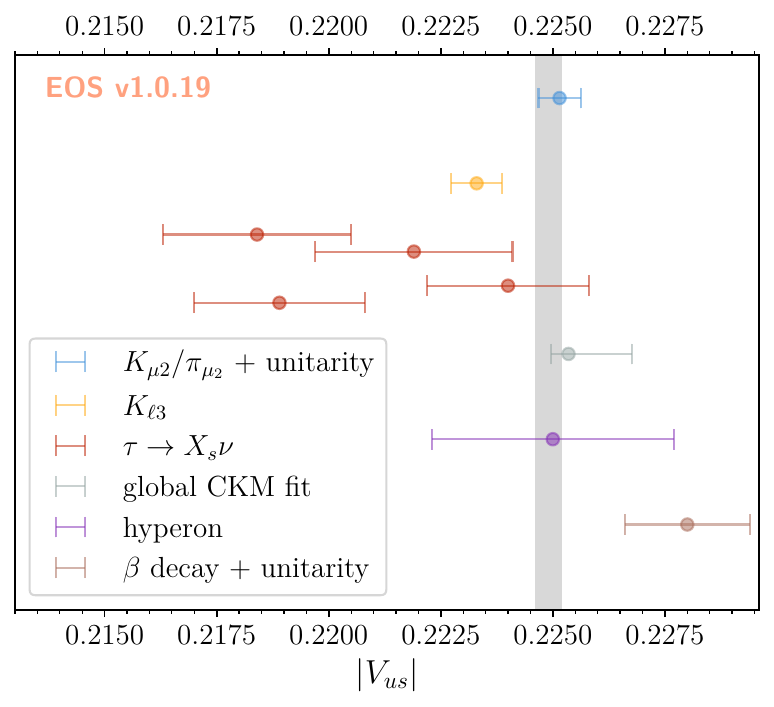}
    \caption{%
        Our result for $\absV{us}$ shown as the shaded vertical band, in comparison to results in the literature.
        We show four independent results obtained from inclusive \Ptau decays, corresponding
        to results from Refs.~\cite{Gamiz:2002nu,Gamiz:2004ar} (top-most data point), Refs.~\cite{Maltman:2015xwa,Maltman:2019xeh} (second from top), Refs.~\cite{RBC:2018uyk,Maltman:2019xeh} (second from bottom), and Ref.~\cite{ExtendedTwistedMass:2024myu} (bottom).
        We refer the interested reader to discussions in the HFLAV report~\cite{HeavyFlavorAveragingGroupHFLAV:2024ctg,HFLAV:tau-vus-web-report} for details of the differences between these determinations.
        The \Kl3, hyperon, and $\beta$ decay + unitarity results are taken from Refs.~\cite{ParticleDataGroup:2024cfk,PDG:Vud_Vus_review}, while the global CKM fit is obtained
        by the CKMfitter collaboration~\cite{Charles:2004jd,CKMfitter:summer2023}.
    }
    \label{fig:ckm:vus}
\end{figure}

Using the representative models given in \cref{eq:results:hmes:metaparameters},
we perform a series of new fits using the CKM fit model and the \texttt{all} data set to determine our CKM posterior.
Having found good agreement with the world average of lattice QCD predictions for $f_+(0)$ in our \textbf{FF} fit,
we now include this input as a prior going forward.
In this way, we determine \absV{us} within a global
analysis. We find that all fits continue to have acceptable $p$ values $\geq \qty{3}{\percent}$ and
therefore provide the first global result for \absV{us} of
\begin{equation}
    \label{eq:results:Vus}
    \absV{us} = 0.2249 \pm 0.0004 (\textrm{stat}) \pm 0.0003 (\textrm{model})\,.
\end{equation}
We show this value in \cref{fig:ckm:vus}, alongside other determinations of \V{us}, and as can been seen our result is in good agreement with previous results.
However, there is still a tension with CKM unitarity.
If we combine our result for \absV{us} with the \absV{ud} result obtained from $\beta$ decay, we find
\begin{equation}
\label{eq:first-row-unitarity-us}
    |\V{ud}|^2 + |\V{us}|^2 = 0.9986 \pm 0.0006 (\V{ud}) \pm 0.0002 (\V{us}) \,,
\end{equation}
showing a continued tension with unitarity of more than \qty{2}{\sigma}.
However, we caution that the small error quoted in \cref{eq:results:Vus} does not yet
account for long-distance electromagnetic effects or isospin breaking effects that can
reach the percent-level.
To improve upon this first global \absV{us} analysis, it is therefore paramount to develop
a framework to include these effects consistently across all of the decays used here.\\

\begin{figure}
    \includegraphics[width=0.9\linewidth]{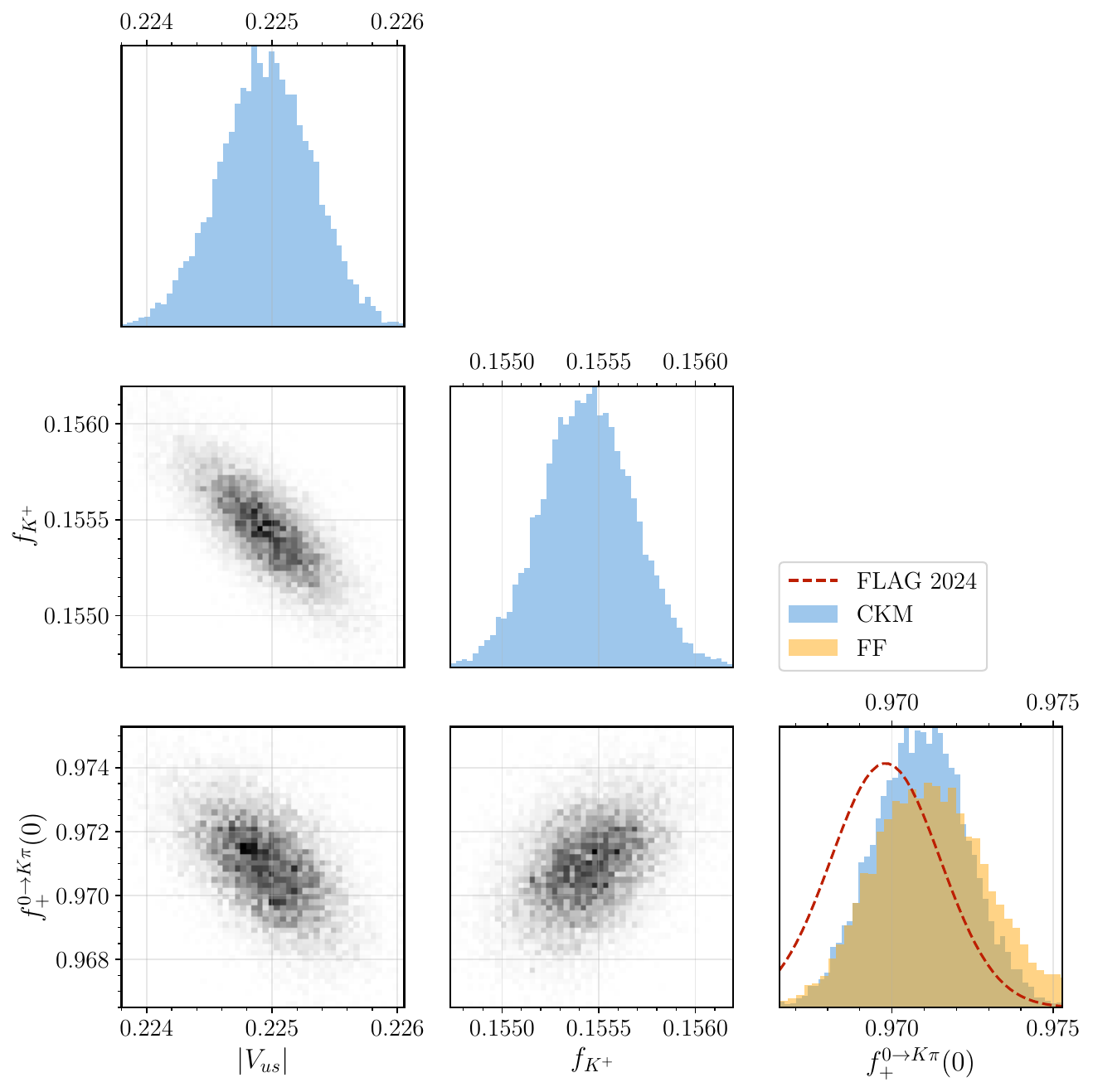}
    \caption{Joint distribution of $\absV{us}$, $f_{K}$, and $f_+(0)$ in our CKM fit model (blue), along with a comparison to the $f_+(0)$ posterior in our FF fit model (orange) and the FLAG 2024 average (red dotted).}
    \label{fig:ckm:correlations}
\end{figure}

Despite now including lattice data on $f_+(0)$ in this fit, our posterior prediction for this quantity remains very similar to the results obtained for the
FF posterior discussed in \cref{sec:results:HME}.
To illustrate this, we show the joint distributions of \absV{us}, $f_{K}$, and $f_+(0)$ in \cref{fig:ckm:correlations}.
We see a strong negative correlation between \absV{us} and both $f_{K}$ and $f_+(0)$.
The correlation between \absV{us} and $f_K$ is expected since the extremely precise $K \to \mu \nu$ branching ratio measurement depends on the product.
The correlation between \absV{us} and $f_+(0)$ can also be understood, since the \Kl3 decays effectively depend on this normalisation along with some mild $q^2$ dependence.
The value obtained for $f_+(0)$ in the CKM posterior remains in good agreement with the FLAG world average~\cite{FlavourLatticeAveragingGroupFLAG:2024oxs}.

\subsection{Fits Searching for BSM Effects}

Given the long-standing discrepancy between \V{us} values extracted from 2-body and 3-body kaon decays, we
use our final fit model to study whether the fit quality is improved by BSM effects.
The BSM model allows for the presence of a right handed current in the form of a non-zero contribution by
the WET Wilson coefficient \CVR (corresponding to the operator defined in \cref{eq:intro:wet:OVR}).
As discussed in \cref{sec:setup:priors}, no further hadronic matrix elements beyond those described in \cref{sec:hmes:preliminaries} arise in this setting.
Our fit gives 
\begin{equation}
    \CVR = \num[uncertainty-descriptors = {stat, model}]{0.0027 \pm 0.0017 \pm 0.0020} \,,
\end{equation}
showing a very mild preference for BSM physics (albeit only at the \qty{1}{\sigma} level due to the model variation we observe), in general agreement with other previous
results~\cite{Grossman:2019bzp,Cirigliano:2021yto,Cirigliano:2022yyo,Crivellin:2022rhw,Cirigliano:2023nol}.
However, again we caveat this statement with the fact that, given our neglect of long distance EM and isospin breaking corrections, our results should not be directly compared with the more sophisticated analyses available in the literature.

\section{Conclusions}
\label{sec:conclusions}

In this paper, we have presented a pilot study showing that data on $K\to \ell\nu$, $K \to \pi \ell\nu$ and $\tau \to K \pi \nu$ decays can be simultaneously described within a single framework, using a new simple form factor parameterisation recently developed for the pion vector form factor.
For this, we make use of theoretical constraints from unitarity in the form of dispersive bounds and chiral perturbation theory calculations for the Callan-Treiman relations, and lattice QCD calculations for the pion and kaon decay constants.
This parametrisation transparently takes into account the resonances present in the \Ptau decay spectrum,
obeys the required
\cref{prop:hmes:properties:no-bound-states,prop:hmes:properties:finite-threshold,prop:hmes:properties:angular-momentum-threshold,prop:hmes:properties:real-below-threshold,prop:hmes:properties:above-threshold-resonances,prop:hmes:properties:large-momentum-limit},
and does not require input on the hadronic phase shift as needed in parameterisations
based on the \Omnes representation.

As discussed in \cref{sec:results:HME}, we find that we are consistently able to fit the \Kresonance{892} resonance pole parameters within our setup and with the current experimental data sets on \Ptau decays.
We find that our results for the pole parameters are in good agreement with the PDG world average.
On the other hand, higher resonances in the $f_+$ spectrum, along with all resonances in the $f_0$ spectrum, are resistant to being fitted.
We attribute these issues to different effects. For the \Kresonance{1680}, we suspect insufficient statistical power due to limited
precision of the data set close to the endpoint of the phase space.
For the \Kresonance{1410} and \Kzeroresonance{1430}, the lack of data on the angular distribution in the $K\pi$ helicity angle
prohibits us from separating both resonances.
We arrive at a nominal fit model that contains the \Kresonance{892}, \Kresonance{1410}, \Kzeroresonance{700}, and \Kzeroresonance{1430}, with the model uncertainty estimated by comparison with a fit model that additionally includes the \Kresonance{1680}, a fit model that increases the $z$ truncation order by one, and a fit model that does both.

In our initial fits, when studying only the hadronic parameters, we further produce posterior predictions for the form factors and their derivatives at $q^2 = 0$.
Since we do not make use of any lattice data on the $K\pi$ form factors, we are able
to use these predictions to test the consistency of our setup.
We do so by comparing them to lattice QCD results of the form factor and experimental results for the derivatives of the form factors. In both cases, we find excellent agreement.
We additionally provide a new estimate for the absolute value of the $f_0$ form factor at $q^2 = m_D^2$ (\cref{eq:results:Ddecay_f0}), which is an important input to non-leptonic $D$ decay calculations, at approximately twice the precision obtained previously in the literature.

With our form factor fits in hand, we then include $\tau \to K \nu$ and $K \to \mu \nu$ experimental data together with the lattice QCD world average of $f_+(0)$ to perform the first global \absV{us} fit, finding a result (\cref{eq:results:Vus}) that is in good agreement with previous individual \absV{us} determinations while still in tension with ``first-row unitarity'' at the level of around \qty{2}{\sigma}.

As a final investigation, we study a model that allows for a Beyond the Standard Model contribution to a right-handed current, corresponding to the operator defined in \cref{eq:intro:wet:OVR}.
Such an effect has been studied in the literature as a possible explanation of the discrepancy seen between individual \absV{us} determinations from 2-body and 3-body decays.
Our fit favours a non-zero central value for the BSM Wilson coefficient $\CVR \simeq 0.0027$, which aligns well with BSM studies
in the literature.
However, given our model uncertainties, our results for a right-handed BSM coupling are compatible with zero at the \qty{1}{\sigma} level.

We identify two key issues that limit further progress.
First, we lack a consistent description of the long-distance electromagnetic corrections to the $K \pi$ matrix elements
both in the semileptonic and in the pair production region
that is independent of the choice of form factor parametrisation.
Second, we lack measurements on the angular distribution in the $K\pi$ helicity angle for the \Ptau decay data.
Such measurements would make it possible to better separate contributions to the $f_+$ and $f_0$ form factors.
We look forward to updating our analysis to provide a more accurate global \absV{us} determination once these
issues are overcome.

\acknowledgments

We are grateful to Stefan Schacht and M\'eril Reboud for useful discussions and suggestions.
We acknowledge support by the Science and Technology Facilities Council through the grants ST/V003941/1 and ST/X003167/1.
MK would like to thank Javier Fuentes Martín and the University of Granada for hospitality provided during the completion of this paper.

\bibliographystyle{JHEP}
\bibliography{refs}

@article{Lepage:1979zb,
    author = "Lepage, G. Peter and Brodsky, Stanley J.",
    title = "{Exclusive Processes in Quantum Chromodynamics: Evolution Equations for Hadronic Wave Functions and the Form-Factors of Mesons}",
    reportNumber = "SLAC-PUB-2343",
    doi = "10.1016/0370-2693(79)90554-9",
    journal = "Phys. Lett. B",
    volume = "87",
    pages = "359--365",
    year = "1979"
}

@article{Muller:2015lua,
    author = {M{\"u}ller, Sarah and Nierste, Ulrich and Schacht, Stefan},
    title = "{Topological amplitudes in $D$ decays to two pseudoscalars: A global analysis with linear $SU(3)_F$ breaking}",
    eprint = "1503.06759",
    archivePrefix = "arXiv",
    primaryClass = "hep-ph",
    reportNumber = "TTP15-015",
    doi = "10.1103/PhysRevD.92.014004",
    journal = "Phys. Rev. D",
    volume = "92",
    number = "1",
    pages = "014004",
    year = "2015"
}

@article{Hill:2006bq,
    author = "Hill, Richard J.",
    title = "{Constraints on the form factors for $K \to \pi \ell \nu$ and implications for $|V_{us}|$}",
    eprint = "hep-ph/0607108",
    archivePrefix = "arXiv",
    reportNumber = "FERMILAB-PUB-06-182-T",
    doi = "10.1103/PhysRevD.74.096006",
    journal = "Phys. Rev. D",
    volume = "74",
    pages = "096006",
    year = "2006"
}

@article{Callan:1966hu,
    author = "Callan, C. G. and Treiman, S. B.",
    title = "{Equal Time Commutators and K Meson Decays}",
    doi = "10.1103/PhysRevLett.16.153",
    journal = "Phys. Rev. Lett.",
    volume = "16",
    pages = "153--157",
    year = "1966"
}

@article{Gasser:1984ux,
    author = "Gasser, J. and Leutwyler, H.",
    title = "{Low-Energy Expansion of Meson Form-Factors}",
    reportNumber = "CERN-TH-3829/84",
    doi = "10.1016/0550-3213(85)90493-6",
    journal = "Nucl. Phys. B",
    volume = "250",
    pages = "517--538",
    year = "1985"
}

@article{Bordone:2019guc,
    author = "Bordone, Marzia and Gubernari, Nico and van Dyk, Danny and Jung, Martin",
    title = "{Heavy-Quark expansion for ${{\bar{B}}_s\rightarrow D^{(*)}_s}$ form factors and unitarity bounds beyond the ${SU(3)_F}$ limit}",
    eprint = "1912.09335",
    archivePrefix = "arXiv",
    primaryClass = "hep-ph",
    reportNumber = "EOS-2019-04, P3H-19-050, SI-HEP-2019-20, TUM-HEP 1241/19",
    doi = "10.1140/epjc/s10052-020-7850-9",
    journal = "Eur. Phys. J. C",
    volume = "80",
    number = "4",
    pages = "347",
    year = "2020"
}

@article{Belle:2007goc,
    author = "Epifanov, D. and others",
    collaboration = "Belle",
    title = "{Study of $\tau^- \to K_{S}^0 \pi^- \nu_{\tau}$ decay at Belle}",
    eprint = "0706.2231",
    archivePrefix = "arXiv",
    primaryClass = "hep-ex",
    reportNumber = "BELLE-PREPRINT-2007-28, KEK-PREPRINT-2007-17",
    doi = "10.1016/j.physletb.2007.08.045",
    journal = "Phys. Lett. B",
    volume = "654",
    pages = "65--73",
    year = "2007"
}

@article{ParticleDataGroup:2024cfk,
    author = "Navas, S. and others",
    collaboration = "Particle Data Group",
    title = "{Review of particle physics}",
    doi = "10.1103/PhysRevD.110.030001",
    journal = "Phys. Rev. D",
    volume = "110",
    number = "3",
    pages = "030001",
    year = "2024"
}

@article{Dowdall:2013rya,
    author = "Dowdall, R. J. and Davies, C. T. H. and Lepage, G. P. and McNeile, C.",
    title = "{$V_{us}$ from $\pi$ and $K$ decay constants in full lattice QCD with physical $u$, $d$, $s$ and $c$ quarks}",
    eprint = "1303.1670",
    archivePrefix = "arXiv",
    primaryClass = "hep-lat",
    doi = "10.1103/PhysRevD.88.074504",
    journal = "Phys. Rev. D",
    volume = "88",
    pages = "074504",
    year = "2013"
}

@article{Carrasco:2014poa,
    author = "Carrasco, N. and others",
    title = "{Leptonic decay constants $f_{K},f_{D},$ and $f_{{D}_{s}}$ with $N_{f} = 2+1+1$ twisted-mass lattice QCD}",
    eprint = "1411.7908",
    archivePrefix = "arXiv",
    primaryClass = "hep-lat",
    doi = "10.1103/PhysRevD.91.054507",
    journal = "Phys. Rev. D",
    volume = "91",
    number = "5",
    pages = "054507",
    year = "2015"
}

@article{FlavourLatticeAveragingGroupFLAG:2024oxs,
    author = "Aoki, Y. and others",
    collaboration = "Flavour Lattice Averaging Group (FLAG)",
    title = "{FLAG review 2024}",
    eprint = "2411.04268",
    archivePrefix = "arXiv",
    primaryClass = "hep-lat",
    reportNumber = "CERN-TH-2024-192, FERMILAB-PUB-24-0785-T",
    doi = "10.1103/nfzp-p5dn",
    journal = "Phys. Rev. D",
    volume = "113",
    number = "1",
    pages = "014508",
    year = "2026"
}

@article{ExtendedTwistedMass:2021qui,
    author = "Alexandrou, C. and others",
    collaboration = "Extended Twisted Mass",
    title = "{Ratio of kaon and pion leptonic decay constants with Nf=2+1+1 Wilson-clover twisted-mass fermions}",
    eprint = "2104.06747",
    archivePrefix = "arXiv",
    primaryClass = "hep-lat",
    doi = "10.1103/PhysRevD.104.074520",
    journal = "Phys. Rev. D",
    volume = "104",
    number = "7",
    pages = "074520",
    year = "2021"
}

@article{Oehme:1966qaf,
    author = "Oehme, Reinhard",
    title = "{Current Algebras and the Suppression of Leptonic Meson Decays with $\Delta S =1$}",
    doi = "10.1103/PhysRevLett.16.215",
    journal = "Phys. Rev. Lett.",
    volume = "16",
    number = "5",
    pages = "215--217",
    year = "1966"
}

@article{Bazavov:2014wgs,
      author         = "{[FNAL/MILC 14A] A. Bazavov} and others",
      title          = "{Charmed and light pseudoscalar meson decay constants
                        from four-flavor lattice QCD with physical light quarks}",
      journal        = "Phys.Rev.",
      number         = "7",
      volume         = "D90",
      pages          = "074509",
      doi            = "10.1103/PhysRevD.90.074509",
      year           = "2014",
      eprint         = "1407.3772",
      archivePrefix  = "arXiv",
      primaryClass   = "hep-lat",
      reportNumber   = "FERMILAB-PUB-14-230-T",
      SLACcitation   = "%%CITATION = ARXIV:1407.3772;%%",
}

@article{Follana:2007uv,
    author = "Follana, E. and Davies, C. T. H. and Lepage, G. P. and Shigemitsu, J.",
    collaboration = "HPQCD, UKQCD",
    title = "{High Precision determination of the $\pi$, $K$, $D$ and $D_{s}$ decay constants from lattice QCD}",
    eprint = "0706.1726",
    archivePrefix = "arXiv",
    primaryClass = "hep-lat",
    doi = "10.1103/PhysRevLett.100.062002",
    journal = "Phys. Rev. Lett.",
    volume = "100",
    pages = "062002",
    year = "2008"
}

@article{MILC:2010hzw,
    author = "Bazavov, A. and others",
    editor = "Rossi, Giancarlo",
    collaboration = "MILC",
    title = "{Results for light pseudoscalar mesons}",
    eprint = "1012.0868",
    archivePrefix = "arXiv",
    primaryClass = "hep-lat",
    doi = "10.22323/1.105.0074",
    journal = "PoS",
    volume = "LATTICE2010",
    pages = "074",
    year = "2010"
}

@article{RBC:2014ntl,
    author = "Blum, T. and others",
    collaboration = "RBC, UKQCD",
    title = "{Domain wall QCD with physical quark masses}",
    eprint = "1411.7017",
    archivePrefix = "arXiv",
    primaryClass = "hep-lat",
    reportNumber = "KEK-TH-1769, RBRC-1095, DAMTP-2014-86",
    doi = "10.1103/PhysRevD.93.074505",
    journal = "Phys. Rev. D",
    volume = "93",
    number = "7",
    pages = "074505",
    year = "2016"
}

@article{CLQCD:2023sdb,
    author = "Hu, Zhi-Cheng and others",
    collaboration = "CLQCD",
    title = "{Quark masses and low-energy constants in the continuum from the tadpole-improved clover ensembles}",
    eprint = "2310.00814",
    archivePrefix = "arXiv",
    primaryClass = "hep-lat",
    doi = "10.1103/PhysRevD.109.054507",
    journal = "Phys. Rev. D",
    volume = "109",
    number = "5",
    pages = "054507",
    year = "2024"
}

@article{KLOE:2005xes,
    author = "Ambrosino, F. and others",
    collaboration = "KLOE",
    title = "{Measurement of the absolute branching ratio for the $K^+ \to \mu^+ \nu(\gamma)$ decay with the KLOE detector}",
    eprint = "hep-ex/0509045",
    archivePrefix = "arXiv",
    doi = "10.1016/j.physletb.2005.11.008",
    journal = "Phys. Lett. B",
    volume = "632",
    pages = "76--80",
    year = "2006"
}

@article{Chiang:1972rp,
    author = "Chiang, I. H. and Rosen, J. L. and Shapiro, S. and Handler, R. and Olsen, S. and Pondrom, L.",
    title = "{$K^+$ Decay in Flight}",
    doi = "10.1103/PhysRevD.6.1254",
    journal = "Phys. Rev. D",
    volume = "6",
    pages = "1254",
    year = "1972"
}

@article{KLOE:2007wlh,
    author = "Ambrosino, F. and others",
    collaboration = "KLOE",
    title = "{Measurement of the charged kaon lifetime with the KLOE detector}",
    eprint = "0712.1112",
    archivePrefix = "arXiv",
    primaryClass = "hep-ex",
    doi = "10.1088/1126-6708/2008/01/073",
    journal = "JHEP",
    volume = "01",
    pages = "073",
    year = "2008"
}

@article{KLOE-2:2019rev,
    author = "Babusci, D. and others",
    collaboration = "KLOE-2",
    title = "{Measurement of the branching fraction for the decay $K_S \to \pi \mu \nu$ with the KLOE detector}",
    eprint = "1912.05990",
    archivePrefix = "arXiv",
    primaryClass = "hep-ex",
    doi = "10.1016/j.physletb.2020.135378",
    journal = "Phys. Lett. B",
    volume = "804",
    pages = "135378",
    year = "2020"
}

@article{KLOE:2002lao,
    author = "Aloisio, A. and others",
    collaboration = "KLOE",
    title = "{Measurement of the branching fraction for the decay $K_{S} \to \pi e \nu$}",
    eprint = "hep-ph/0203232",
    archivePrefix = "arXiv",
    reportNumber = "LNF-02-001",
    doi = "10.1016/S0370-2693(02)01733-1",
    journal = "Phys. Lett. B",
    volume = "535",
    pages = "37--42",
    year = "2002"
}

@article{KTeV:2004hpx,
    author = "Alexopoulos, T. and others",
    collaboration = "KTeV",
    title = "{Measurements of $K_L$ Branching Fractions and the CP Violation Parameter $|\eta_{\pm}|$}",
    eprint = "hep-ex/0406002",
    archivePrefix = "arXiv",
    reportNumber = "FERMILAB-PUB-04-165-E",
    doi = "10.1103/PhysRevD.70.092006",
    journal = "Phys. Rev. D",
    volume = "70",
    pages = "092006",
    year = "2004"
}

@article{KLOE:2005vdt,
    author = "Ambrosino, F. and others",
    collaboration = "KLOE",
    title = "{Measurements of the absolute branching ratios for the dominant $K_L$ decays, the $K_L$ lifetime, and $V_{us}$ with the KLOE detector}",
    eprint = "hep-ex/0508027",
    archivePrefix = "arXiv",
    doi = "10.1016/j.physletb.2005.10.018",
    journal = "Phys. Lett. B",
    volume = "632",
    pages = "43--50",
    year = "2006"
}

@article{CLEO:1994bvq,
    author = "Battle, M. and others",
    collaboration = "CLEO",
    title = "{Measurement of Cabibbo suppressed decays of the $\tau$ lepton}",
    eprint = "hep-ph/9403329",
    archivePrefix = "arXiv",
    reportNumber = "CLNS-94-1273, CLEO-94-8, CLNS94-1273",
    doi = "10.1103/PhysRevLett.73.1079",
    journal = "Phys. Rev. Lett.",
    volume = "73",
    pages = "1079--1083",
    year = "1994"
}

@article{DELPHI:1994gzt,
    author = "Abreu, P. and others",
    collaboration = "DELPHI",
    title = "{Charged kaon production in $\tau$ decays at LEP}",
    reportNumber = "CERN-PPE-94-088, CERN-PPE-94-88",
    doi = "10.1016/0370-2693(94)90711-0",
    journal = "Phys. Lett. B",
    volume = "334",
    pages = "435--449",
    year = "1994"
}

@article{ALEPH:1999jxs,
    author = "Barate, R. and others",
    collaboration = "ALEPH",
    title = "{One prong $\tau$ decays with kaons}",
    eprint = "hep-ex/9903014",
    archivePrefix = "arXiv",
    reportNumber = "CERN-EP-99-025",
    doi = "10.1007/s100529900146",
    journal = "Eur. Phys. J. C",
    volume = "10",
    pages = "1--18",
    year = "1999"
}

@article{OPAL:2000fde,
    author = "Abbiendi, G. and others",
    collaboration = "OPAL",
    title = "{A Study of one prong $\tau$ decays with a charged kaon}",
    eprint = "hep-ex/0009017",
    archivePrefix = "arXiv",
    reportNumber = "CERN-EP-2000-091, OPAL-PR-315",
    doi = "10.1007/s100520100632",
    journal = "Eur. Phys. J. C",
    volume = "19",
    pages = "653--665",
    year = "2001"
}

@article{Ling-Fong:1971tjw,
    author = "Ling-Fong, Li. and Pagels, H.",
    title = "{Rigorous bound on $K_{\ell 3}$ decay amplitudes}",
    doi = "10.1103/PhysRevD.3.2191",
    journal = "Phys. Rev. D",
    volume = "3",
    pages = "2191--2195",
    year = "1971"
}

@article{Okubo:1971my,
    author = "Okubo, S.",
    title = "{New improved bounds for $K_{\ell 3}$ parameters}",
    doi = "10.1103/PhysRevD.4.725",
    journal = "Phys. Rev. D",
    volume = "4",
    pages = "725--733",
    year = "1971"
}

@article{Kirk:2024oyl,
    author = "Kirk, Matthew and Kubis, Bastian and Reboud, M{\'e}ril and van Dyk, Danny",
    title = "{A simple parametrisation of the pion form factor}",
    eprint = "2410.13764",
    archivePrefix = "arXiv",
    primaryClass = "hep-ph",
    reportNumber = "EOS-2024-03, IPPP/24/65",
    doi = "10.1016/j.physletb.2025.139266",
    journal = "Phys. Lett. B",
    volume = "861",
    pages = "139266",
    year = "2025"
}

@article{Bourrely:2008za,
    author = "Bourrely, Claude and Caprini, Irinel and Lellouch, Laurent",
    title = "{Model-independent description of $B \to \pi \ell \nu$ decays and a determination of $|V_{ub}|$}",
    eprint = "0807.2722",
    archivePrefix = "arXiv",
    primaryClass = "hep-ph",
    reportNumber = "CPT-P36-2007",
    doi = "10.1103/PhysRevD.82.099902",
    journal = "Phys. Rev. D",
    volume = "79",
    pages = "013008",
    year = "2009",
    note = "[Erratum: Phys.Rev.D 82, 099902 (2010)]"
}

@article{Becher:2005bg,
    author = "Becher, Thomas and Hill, Richard J.",
    title = "{Comment on form-factor shape and extraction of $|V_{ub}|$ from $B \to \pi \ell \nu$}",
    eprint = "hep-ph/0509090",
    archivePrefix = "arXiv",
    reportNumber = "FERMILAB-PUB-05-385-T, SLAC-PUB-11468",
    doi = "10.1016/j.physletb.2005.11.063",
    journal = "Phys. Lett. B",
    volume = "633",
    pages = "61--69",
    year = "2006"
}

@article{Buck:1998kp,
    author = "Buck, W. W. and Lebed, Richard F.",
    title = "{New constraints on dispersive form-factor parameterizations from the timelike region}",
    eprint = "hep-ph/9802369",
    archivePrefix = "arXiv",
    reportNumber = "JLAB-THY-98-04",
    doi = "10.1103/PhysRevD.58.056001",
    journal = "Phys. Rev. D",
    volume = "58",
    pages = "056001",
    year = "1998"
}

@article{Higson:2018,
	doi = {10.1007/s11222-018-9844-0},
	url = {https://doi.org/10.1007%2Fs11222-018-9844-0},
	year = 2018,
	month = {dec},
	publisher = {Springer Science and Business Media {LLC}},
	volume = {29},
	number = {5},
	pages = {891--913},
	author = {Edward Higson and Will Handley and Michael Hobson and Anthony Lasenby},
	title = {Dynamic nested sampling: an improved algorithm for parameter estimation and evidence calculation},
	journal = {Statistics and Computing}
}

@misc{dynesty:v2.0.3,
  author       = {Sergey Koposov and
                  Josh Speagle and
                  Kyle Barbary and
                  Gregory Ashton and
                  Ed Bennett and
                  Johannes Buchner and
                  Carl Scheffler and
                  Ben Cook and
                  Colm Talbot and
                  James Guillochon and
                  Patricio Cubillos and
                  Andrés Asensio Ramos and
                  Ben Johnson and
                  Dustin Lang and
                  Ilya and
                  Matthieu Dartiailh and
                  Alex Nitz and
                  Andrew McCluskey and
                  Anne Archibald and
                  Christoph Deil and
                  Dan Foreman-Mackey and
                  Danny Goldstein and
                  Erik Tollerud and
                  Joel Leja and
                  Matthew Kirk and
                  Matt Pitkin and
                  Patrick Sheehan and
                  Phillip Cargile and
                  ruskin23 and
                  Ruth Angus},
  title        = {\texttt{dynesty} version 2.0.3},
  month        = dec,
  year         = 2022,
  publisher    = {Zenodo},
  version      = {v2.0.3},
  doi          = {\href{https://doi.org/10.5281/zenodo.7388523}{10.5281/zenodo.7388523}},
  url          = {https://doi.org/10.5281/zenodo.7388523}
}

@software{EOS:v1.0.19,
  author       = {Danny van Dyk and
                  Méril Reboud and
                  Frederik Beaujean and
                  Matthew Kirk and
                  Nico Gubernari and
                  Florian Herren and
                  Philip Lüghausen and
                  Domagoj Leljak and
                  Stefan Meiser and
                  Stephan Kürten and
                  Carolina Bolognani and
                  Lorenz Gaertner and
                  Viktor Kuschke and
                  Filip Novak and
                  Christoph Bobeth and
                  Dominik Suelmann and
                  Elena Graverini and
                  Ery McPartland and
                  Marta Burgos and
                  Martin Ritter and
                  Charles Earnshaw and
                  Thomas Blake and
                  Marzia Bordone and
                  Eike Eberhard and
                  Nienke C. Balz and
                  K. Keri Vos and
                  Ismo Toijala and
                  Ahmet Kokulu and
                  Eduardo Romero and
                  Romy O'Connor and
                  Maximilian Hoverath and
                  Jonas Eschle},
  title        = {\texttt{EOS} version 1.0.19},
  month        = dec,
  year         = 2025,
  publisher    = {Zenodo},
  version      = {v1.0.19},
  doi          = {\href{https://doi.org/10.5281/zenodo.17792609}{10.5281/zenodo.17792609}},
  url          = {https://doi.org/10.5281/zenodo.17792609},
  swhid        = {swh:1:dir:15380be12f8d8cbbcae967f3f24092d4036982e1
                   ;origin=https://doi.org/10.5281/zenodo.3376590;vis
                   it=swh:1:snp:47012bd580b65b25da243a25580331b307919
                   7f5;anchor=swh:1:rel:ff533a1f851aab5c39eca084e35bd
                   f688aa28e6c;path=eos-eos-37907e2
                  },
}

@dataset{EOS-DATA-2025-05,
  author       = "Kirk, Matthew and van Dyk, Danny",
  title        = "{\texttt{EOS/DATA-2025-05}: Supplementary material for 
                   \texttt{EOS/ANALYSIS-2025-01}}",
  month        = nov,
  year         = 2025,
  publisher    = {Zenodo},
  doi          = {\href{https://doi.org/10.5281/zenodo.18788354}{10.5281/zenodo.18788354}},
}

@article{EOSAuthors:2021xpv,
    author = "van Dyk, Danny and others",
    collaboration = "EOS Authors",
    title = "{\EOS: a software for flavor physics phenomenology}",
    eprint = "2111.15428",
    archivePrefix = "arXiv",
    primaryClass = "hep-ph",
    reportNumber = "EOS-2021-04, TUM-HEP 1371/21, P3H-21-094, SI-HEP-2021-32",
    doi = "10.1140/epjc/s10052-022-10177-4",
    journal = "Eur. Phys. J. C",
    volume = "82",
    number = "6",
    pages = "569",
    year = "2022"
}

@article{Bernard:2009zm,
    author = "Bernard, Veronique and Oertel, Micaela and Passemar, Emilie and Stern, Jan",
    title = "{Dispersive representation and shape of the $K_{\ell 3}$ form factors: Robustness}",
    eprint = "0903.1654",
    archivePrefix = "arXiv",
    primaryClass = "hep-ph",
    doi = "10.1103/PhysRevD.80.034034",
    journal = "Phys. Rev. D",
    volume = "80",
    pages = "034034",
    year = "2009"
}

@article{Dashen:1969bh,
    author = "Dashen, Roger F. and Weinstein, M.",
    title = "{Theorem on the form-factors in $K_{\ell 3}$ decay}",
    doi = "10.1103/PhysRevLett.22.1337",
    journal = "Phys. Rev. Lett.",
    volume = "22",
    pages = "1337--1340",
    year = "1969"
}

@article{Boyd:1995cf,
    author = "Boyd, C. Glenn and Grinstein, Benjamin and Lebed, Richard F.",
    title = "{Model independent extraction of $|V_{cb}|$ using dispersion relations}",
    eprint = "hep-ph/9504235",
    archivePrefix = "arXiv",
    reportNumber = "UCSD-PTH-95-03",
    doi = "10.1016/0370-2693(95)00480-9",
    journal = "Phys. Lett. B",
    volume = "353",
    pages = "306--312",
    year = "1995"
}

@article{Caprini:1997mu,
    author = "Caprini, Irinel and Lellouch, Laurent and Neubert, Matthias",
    title = "{Dispersive bounds on the shape of $\bar{B} \to D^{(*)} \ell \bar\nu$ form-factors}",
    eprint = "hep-ph/9712417",
    archivePrefix = "arXiv",
    reportNumber = "CERN-TH-97-091, CPT-97-P3480",
    doi = "10.1016/S0550-3213(98)00350-2",
    journal = "Nucl. Phys. B",
    volume = "530",
    pages = "153--181",
    year = "1998"
}

@article{Belle:2014mfl,
    author = "Ryu, S. and others",
    collaboration = "Belle",
    title = "{Measurements of Branching Fractions of $\tau$ Lepton Decays with one or more $K^{0}_{S}$}",
    eprint = "1402.5213",
    archivePrefix = "arXiv",
    primaryClass = "hep-ex",
    doi = "10.1103/PhysRevD.89.072009",
    journal = "Phys. Rev. D",
    volume = "89",
    number = "7",
    pages = "072009",
    year = "2014"
}

@article{OPAL:1999bbs,
    author = "Abbiendi, G. and others",
    collaboration = "OPAL",
    title = "{$\tau$ decays with neutral kaons}",
    eprint = "hep-ex/9911029",
    archivePrefix = "arXiv",
    reportNumber = "CERN-EP-99-154",
    doi = "10.1007/s100520000317",
    journal = "Eur. Phys. J. C",
    volume = "13",
    pages = "213--223",
    year = "2000"
}

@article{L3:1995cos,
    author = "Acciarri, M. and others",
    collaboration = "L3",
    title = "{One prong $\tau$ decays with neutral kaons}",
    reportNumber = "CERN-PPE-95-42, CERN-PPE-95-042",
    doi = "10.1016/0370-2693(95)00509-J",
    journal = "Phys. Lett. B",
    volume = "352",
    pages = "487--497",
    year = "1995"
}

@article{BaBar:2007yir,
    author = "Aubert, Bernard and others",
    collaboration = "BaBar",
    title = "{Measurement of the $\tau^{-} \to K^{-} \pi^0 \nu_{tau}$ branching fraction}",
    eprint = "0707.2922",
    archivePrefix = "arXiv",
    primaryClass = "hep-ex",
    reportNumber = "SLAC-PUB-12681, BABAR-PUB-07-036",
    doi = "10.1103/PhysRevD.76.051104",
    journal = "Phys. Rev. D",
    volume = "76",
    pages = "051104",
    year = "2007"
}

@article{OPAL:2004icu,
    author = "Abbiendi, G. and others",
    collaboration = "OPAL",
    title = "{Measurement of the strange spectral function in hadronic $\tau$ decays}",
    eprint = "hep-ex/0406007",
    archivePrefix = "arXiv",
    reportNumber = "CERN-PH-EP-2004-010",
    doi = "10.1140/epjc/s2004-01877-2",
    journal = "Eur. Phys. J. C",
    volume = "35",
    pages = "437--455",
    year = "2004"
}

@article{FermilabLattice:2018zqv,
    author = "Bazavov, A. and others",
    collaboration = "Fermilab Lattice, MILC",
    title = "{$|V_{us}|$ from $K_{\ell 3}$ decay and four-flavor lattice QCD}",
    eprint = "1809.02827",
    archivePrefix = "arXiv",
    primaryClass = "hep-lat",
    reportNumber = "FERMILAB-PUB-18-439-T",
    doi = "10.1103/PhysRevD.99.114509",
    journal = "Phys. Rev. D",
    volume = "99",
    number = "11",
    pages = "114509",
    year = "2019"
}

@article{Carrasco:2016kpy,
    author = "Carrasco, N. and Lami, P. and Lubicz, V. and Riggio, L. and Simula, S. and Tarantino, C.",
    title = "{$K \to \pi$ semileptonic form factors with $N_f=2+1+1$ twisted mass fermions}",
    eprint = "1602.04113",
    archivePrefix = "arXiv",
    primaryClass = "hep-lat",
    reportNumber = "PREPRINT-RM3-TH-16-2",
    doi = "10.1103/PhysRevD.93.114512",
    journal = "Phys. Rev. D",
    volume = "93",
    number = "11",
    pages = "114512",
    year = "2016"
}

@article{Balz:2025auk,
    author = "Balz, Nienke C. and Herren, Florian and Kubis, Bastian and Mutke, Simon and Reboud, M{\'e}ril",
    title = "{Advanced parametrisations for hadronic form factors}",
    eprint = "2510.25584",
    archivePrefix = "arXiv",
    primaryClass = "hep-ph",
    reportNumber = "ZU-TH 70/25",
    month = "10",
    year = "2025"
}

@article{NA482:2018rgv,
    author = "Batley, John Richard and others",
    collaboration = "NA48/2",
    title = "{Measurement of the form factors of charged kaon semileptonic decays}",
    eprint = "1808.09041",
    archivePrefix = "arXiv",
    primaryClass = "hep-ex",
    reportNumber = "CERN-EP-2018-231",
    doi = "10.1007/JHEP10(2018)150",
    journal = "JHEP",
    volume = "10",
    pages = "150",
    year = "2018"
}

@article{Bordone:2019vic,
    author = "Bordone, Marzia and Jung, Martin and van Dyk, Danny",
    title = "{Theory determination of $\bar{B}\to D^{(*)}\ell^-\bar\nu$ form factors at $\mathcal{O}(1/m_c^2)$}",
    eprint = "1908.09398",
    archivePrefix = "arXiv",
    primaryClass = "hep-ph",
    reportNumber = "EOS-2019-02, P3H-19-020, SI-HEP-2019-08, TUM-HEP 1211/19",
    doi = "10.1140/epjc/s10052-020-7616-4",
    journal = "Eur. Phys. J. C",
    volume = "80",
    number = "2",
    pages = "74",
    year = "2020"
}

@article{Grossman:2019bzp,
    author = "Grossman, Yuval and Passemar, Emilie and Schacht, Stefan",
    title = "{On the Statistical Treatment of the Cabibbo Angle Anomaly}",
    eprint = "1911.07821",
    archivePrefix = "arXiv",
    primaryClass = "hep-ph",
    reportNumber = "JLAB-THY-19-3050",
    doi = "10.1007/JHEP07(2020)068",
    journal = "JHEP",
    volume = "07",
    pages = "068",
    year = "2020"
}

@article{Belfatto:2021jhf,
    author = "Belfatto, Benedetta and Berezhiani, Zurab",
    title = "{Are the CKM anomalies induced by vector-like quarks? Limits from flavor changing and Standard Model precision tests}",
    eprint = "2103.05549",
    archivePrefix = "arXiv",
    primaryClass = "hep-ph",
    doi = "10.1007/JHEP10(2021)079",
    journal = "JHEP",
    volume = "10",
    pages = "079",
    year = "2021"
}

@article{Cirigliano:2022yyo,
    author = "Cirigliano, Vincenzo and Crivellin, Andreas and Hoferichter, Martin and Moulson, Matthew",
    title = "{Scrutinizing CKM unitarity with a new measurement of the K{\ensuremath{\mu}}3/K{\ensuremath{\mu}}2 branching fraction}",
    eprint = "2208.11707",
    archivePrefix = "arXiv",
    primaryClass = "hep-ph",
    reportNumber = "INT-PUB-22-024, PSI-PR-22-28, ZU-TH 43/22",
    doi = "10.1016/j.physletb.2023.137748",
    journal = "Phys. Lett. B",
    volume = "838",
    pages = "137748",
    year = "2023"
}

@article{Cirigliano:2023nol,
    author = "Cirigliano, Vincenzo and Dekens, Wouter and de Vries, Jordy and Mereghetti, Emanuele and Tong, Tom",
    title = "{Anomalies in global SMEFT analyses. A case study of first-row CKM unitarity}",
    eprint = "2311.00021",
    archivePrefix = "arXiv",
    primaryClass = "hep-ph",
    doi = "10.1007/JHEP03(2024)033",
    journal = "JHEP",
    volume = "03",
    pages = "033",
    year = "2024"
}

@article{Crivellin:2022rhw,
    author = "Crivellin, Andreas and Kirk, Matthew and Kitahara, Teppei and Mescia, Federico",
    title = "{Global fit of modified quark couplings to EW gauge bosons and vector-like quarks in light of the Cabibbo angle anomaly}",
    eprint = "2212.06862",
    archivePrefix = "arXiv",
    primaryClass = "hep-ph",
    reportNumber = "PSI-PR-22-37, ZU-TH-61/22, KEK-TH-2480",
    doi = "10.1007/JHEP03(2023)234",
    journal = "JHEP",
    volume = "03",
    pages = "234",
    year = "2023"
}

@article{Cirigliano:2021yto,
    author = "Cirigliano, Vincenzo and D{\'\i}az-Calder{\'o}n, David and Falkowski, Adam and Gonz{\'a}lez-Alonso, Mart{\'\i}n and Rodr{\'\i}guez-S{\'a}nchez, Antonio",
    title = "{Semileptonic $\tau$ decays beyond the Standard Model}",
    eprint = "2112.02087",
    archivePrefix = "arXiv",
    primaryClass = "hep-ph",
    doi = "10.1007/JHEP04(2022)152",
    journal = "JHEP",
    volume = "04",
    pages = "152",
    year = "2022"
}

@article{Belfatto:2023tbv,
    author = "Belfatto, Benedetta and Trifinopoulos, Sokratis",
    title = "{Cabibbo angle anomalies and oblique corrections: The remarkable role of the vectorlike quark doublet}",
    eprint = "2302.14097",
    archivePrefix = "arXiv",
    primaryClass = "hep-ph",
    reportNumber = "MIT-CTP/5538",
    doi = "10.1103/PhysRevD.108.035022",
    journal = "Phys. Rev. D",
    volume = "108",
    number = "3",
    pages = "035022",
    year = "2023"
}

@article{Jamin:2006tk,
    author = "Jamin, Matthias and Pich, Antonio and Portoles, Jorge",
    title = "{Spectral distribution for the decay $\tau^- \to \nu_{\tau} K \pi$}",
    eprint = "hep-ph/0605096",
    archivePrefix = "arXiv",
    reportNumber = "UAB-FT-601, IFIC-06-10, FTUV-06-0509",
    doi = "10.1016/j.physletb.2006.06.058",
    journal = "Phys. Lett. B",
    volume = "640",
    pages = "176--181",
    year = "2006"
}

@article{Jamin:2008qg,
    author = "Jamin, Matthias and Pich, Antonio and Portoles, Jorge",
    title = "{What can be learned from the Belle spectrum for the decay $\tau^- \to \nu_{\tau} K_{S}^0 \pi^-$}",
    eprint = "0803.1786",
    archivePrefix = "arXiv",
    primaryClass = "hep-ph",
    reportNumber = "UAB-FT-642, IFIC-08-16, FTUV-08-0312",
    doi = "10.1016/j.physletb.2008.04.049",
    journal = "Phys. Lett. B",
    volume = "664",
    pages = "78--83",
    year = "2008"
}

@article{Boito:2008fq,
    author = "Boito, Diogo R. and Escribano, Rafel and Jamin, Matthias",
    title = "{$K \pi$ vector form-factor, dispersive constraints and $\tau^- \to \nu_{\tau} K \pi$ decays}",
    eprint = "0807.4883",
    archivePrefix = "arXiv",
    primaryClass = "hep-ph",
    reportNumber = "UAB-FT-653",
    doi = "10.1140/epjc/s10052-008-0834-9",
    journal = "Eur. Phys. J. C",
    volume = "59",
    pages = "821--829",
    year = "2009"
}

@article{Boito:2010me,
    author = "Boito, D. R. and Escribano, R. and Jamin, M.",
    title = "{$K \pi$ vector form factor constrained by $\tau \to K\pi \nu_\tau$ and $K_{\ell 3}$ decays}",
    eprint = "1007.1858",
    archivePrefix = "arXiv",
    primaryClass = "hep-ph",
    reportNumber = "UAB-FT-682",
    doi = "10.1007/JHEP09(2010)031",
    journal = "JHEP",
    volume = "09",
    pages = "031",
    year = "2010"
}

@phdthesis{Rendon:2021nvu,
    author = "Rend{\'o}n, Javier",
    title = "{Exclusive hadronic {\ensuremath{\tau}} decays as probes of non-SM interactions}",
    school = "CINVESTAV, IPN",
    month = "10",
    year = "2021"
}

@article{Gamiz:2004ar,
    author = "Gamiz, Elvira and Jamin, Matthias and Pich, Antonio and Prades, Joaquim and Schwab, Felix",
    title = "{V(us) and m(s) from hadronic tau decays}",
    eprint = "hep-ph/0408044",
    archivePrefix = "arXiv",
    reportNumber = "HD-THEP-04-30, IFIC-04-43, FTUV-04-0803, CAFPE-34-04, UG-FT-164-04, MPP-2004-93, TUM-HEP-555-04",
    doi = "10.1103/PhysRevLett.94.011803",
    journal = "Phys. Rev. Lett.",
    volume = "94",
    pages = "011803",
    year = "2005"
}

@article{Gamiz:2002nu,
    author = "Gamiz, E. and Jamin, M. and Pich, A. and Prades, J. and Schwab, F.",
    title = "{Determination of $m_s$ and $|V_{us}|$ from hadronic $\tau$ decays}",
    eprint = "hep-ph/0212230",
    archivePrefix = "arXiv",
    reportNumber = "IFIC-02-61, CAFPE-14-02, UG-FT-144-02, FTUV-02-1216, HD-THEP-02-42",
    doi = "10.1088/1126-6708/2003/01/060",
    journal = "JHEP",
    volume = "01",
    pages = "060",
    year = "2003"
}

@article{Maltman:2015xwa,
    author = "Maltman, Kim and Hudspith, Renwick J. and Lewis, Randy and Wolfe, Carl E. and Zanotti, James",
    title = "{A resolution of the inclusive flavor-breaking sum rule $\tau$ $V_{us}$ puzzle}",
    eprint = "1510.06954",
    archivePrefix = "arXiv",
    primaryClass = "hep-ph",
    doi = "10.22323/1.251.0260",
    journal = "PoS",
    volume = "LATTICE2015",
    pages = "260",
    year = "2016"
}

@article{Maltman:2019xeh,
    author = "Maltman, Kim and others",
    title = "{Current Status of inclusive hadronic $\tau$ determinations of $|V_{us}|$}",
    doi = "10.21468/SciPostPhysProc.1.006",
    journal = "SciPost Phys. Proc.",
    volume = "1",
    pages = "006",
    year = "2019"
}

@article{RBC:2018uyk,
    author = {Boyle, Peter and Hudspith, Renwick James and Izubuchi, Taku and J{\"u}ttner, Andreas and Lehner, Christoph and Lewis, Randy and Maltman, Kim and Ohki, Hiroshi and Portelli, Antonin and Spraggs, Matthew},
    collaboration = "RBC, UKQCD",
    title = "{Novel $|V_{us}|$ Determination Using Inclusive Strange $\tau$ Decay and Lattice Hadronic Vacuum Polarization Functions}",
    eprint = "1803.07228",
    archivePrefix = "arXiv",
    primaryClass = "hep-lat",
    doi = "10.1103/PhysRevLett.121.202003",
    journal = "Phys. Rev. Lett.",
    volume = "121",
    number = "20",
    pages = "202003",
    year = "2018"
}

@article{HeavyFlavorAveragingGroupHFLAV:2024ctg,
    author = "Banerjee, Sw. and others",
    collaboration = "Heavy Flavor Averaging Group (HFLAV)",
    title = "{Averages of b-hadron, c-hadron, and {\ensuremath{\tau}}-lepton properties as of 2023}",
    eprint = "2411.18639",
    archivePrefix = "arXiv",
    primaryClass = "hep-ex",
    doi = "10.1103/x87q-tld5",
    journal = "Phys. Rev. D",
    volume = "113",
    number = "1",
    pages = "012008",
    year = "2026"
}

@misc{HFLAV:tau-vus-web-report,
    title = "{HFLAV Tau 2023 web report}",
    author = "{Sw. Banerjee, M. Chrząszcz, K. Hayasaka, A. Lusiani, M. Roney, B. Shwartz}",
    url = "https://hflav-eos.web.cern.ch/hflav-eos/tau/end-2023/vus.html"
}

@article{ExtendedTwistedMass:2024myu,
    author = "Alexandrou, Constantia and others",
    collaboration = "Extended Twisted Mass",
    title = "{Inclusive Hadronic Decay Rate of the {\ensuremath{\tau}} Lepton from Lattice QCD: The u{\textasciimacron}s Flavor Channel and the Cabibbo Angle}",
    eprint = "2403.05404",
    archivePrefix = "arXiv",
    primaryClass = "hep-lat",
    doi = "10.1103/PhysRevLett.132.261901",
    journal = "Phys. Rev. Lett.",
    volume = "132",
    number = "26",
    pages = "261901",
    year = "2024"
}

@article{Charles:2004jd,
    author = "Charles, J. and Hocker, Andreas and Lacker, H. and Laplace, S. and Le Diberder, F. R. and Malcles, J. and Ocariz, J. and Pivk, M. and Roos, L.",
    collaboration = "CKMfitter Group",
    title = "{CP violation and the CKM matrix: Assessing the impact of the asymmetric $B$ factories}",
    eprint = "hep-ph/0406184",
    archivePrefix = "arXiv",
    reportNumber = "CPT-2004-P-030, LAL-04-21, LAPP-EXP-2004-01, LPNHE-2004-01",
    doi = "10.1140/epjc/s2005-02169-1",
    journal = "Eur. Phys. J. C",
    volume = "41",
    number = "1",
    pages = "1--131",
    year = "2005"
}

@misc{CKMfitter:summer2023,
    title = "{CKMfitter Summer 2023 update}",
    url = "http://ckmfitter.in2p3.fr/www/results/plots_summer23/num/ckmEval_results_summer23.html"
}

@misc{PDG:Vud_Vus_review,
    collaboration = "{Particle Data Group}",
    title = "{$V_{ud}$, $V_{us}$ the Cabibbo Angle, and CKM Unitarity}",
    url  = "https://pdg.lbl.gov/2025/reviews/rpp2024-rev-vud-vus.pdf"
}

@article{Moretti:2025qxt,
    author = {Moretti, Francesco and Gorbahn, Martin and J{\"a}ger, Sebastian},
    title = "{Beyond Leading Logarithms in $g_V$: The Semileptonic Weak Hamiltonian at $\mathcal{O}(\alpha \, \alpha_s^2)$}",
    eprint = "2510.27648",
    archivePrefix = "arXiv",
    primaryClass = "hep-ph",
    reportNumber = "P3H-25-087, TTP25-041",
    month = "10",
    year = "2025"
}

@article{Cao:2025zxs,
    author = "Cao, Zehua and Hill, Richard J. and Plestid, Ryan and Vander Griend, Peter",
    title = "{The $Z\alpha^2$ correction to superallowed beta decays in effective field theory and implications for $|V_{ud}|$}",
    eprint = "2511.05446",
    archivePrefix = "arXiv",
    primaryClass = "hep-ph",
    reportNumber = "FERMILAB-PUB-25-0773-T, CALT-TH-2025-032, CERN-TH-2025-20",
    month = "11",
    year = "2025"
}

@article{Crosas:2025xyv,
    author = "Crosas, {\`O}scar L. and Mereghetti, Emanuele",
    title = "{Radiative corrections to superallowed beta decays at $ \mathcal{O}\left({\alpha}^2Z\right) $}",
    eprint = "2511.05481",
    archivePrefix = "arXiv",
    primaryClass = "hep-ph",
    reportNumber = "ZU-TH 78/25, LA-UR-25-31083",
    doi = "10.1007/JHEP02(2026)114",
    journal = "JHEP",
    volume = "02",
    pages = "114",
    year = "2026"
}

@article{Gorchtein:2025wli,
    author = "Gorchtein, M. and Katyal, V. and Ohayon, B. and Sahoo, B. K. and Seng, Chien-Yeah",
    title = "{Cabibbo-Kobayashi-Maskawa unitarity deficit reduction via finite nuclear size}",
    eprint = "2502.17070",
    archivePrefix = "arXiv",
    primaryClass = "nucl-th",
    reportNumber = "NT@UW-25-3",
    doi = "10.1103/z8g6-9j25",
    journal = "Phys. Rev. Res.",
    volume = "7",
    number = "4",
    pages = "L042002",
    year = "2025"
}

@article{Aebischer:2017ugx,
    author = "Aebischer, Jason and others",
    title = "{WCxf: an exchange format for Wilson coefficients beyond the Standard Model}",
    eprint = "1712.05298",
    archivePrefix = "arXiv",
    primaryClass = "hep-ph",
    reportNumber = "IFIC-17-61, TUM-HEP-1117-17, LMU-ASC-74-17, IFIC/17-61, KA-TP-38-2017, TUM-HEP-1117/17, LMU-ASC 74/17",
    doi = "10.1016/j.cpc.2018.05.022",
    journal = "Comput. Phys. Commun.",
    volume = "232",
    pages = "71--83",
    year = "2018"
}

@misc{PDG:FF-review,
    collaboration = "{Particle Data Group}",
    title = "{Form Factors for Semileptonic Kaon ($K_{\ell 3}$), Radiative Pion
($\pi_{\ell 2 \gamma}$) and Kaon ($K_{\ell 2 \gamma}$) Decays}",
    url = "https://pdg.lbl.gov/2025/reviews/rpp2024-rev-form-factors-radiative-pik-decays.pdf"
}

@article{Crivellin:2021bkd,
    author = "Crivellin, Andreas and Hoferichter, Martin and Kirk, Matthew and Manzari, Claudio Andrea and Schnell, Luc",
    title = "{First-generation new physics in simplified models: from low-energy parity violation to the LHC}",
    eprint = "2107.13569",
    archivePrefix = "arXiv",
    primaryClass = "hep-ph",
    reportNumber = "CERN-TH-2021-112, PSI-21-16, ZU-TH 32/21",
    doi = "10.1007/JHEP10(2021)221",
    journal = "JHEP",
    volume = "10",
    pages = "221",
    year = "2021"
}

@article{Crivellin:2021njn,
    author = "Crivellin, Andreas and Hoferichter, Martin and Manzari, Claudio Andrea",
    title = "{Fermi Constant from Muon Decay Versus Electroweak Fits and Cabibbo-Kobayashi-Maskawa Unitarity}",
    eprint = "2102.02825",
    archivePrefix = "arXiv",
    primaryClass = "hep-ph",
    reportNumber = "CERN-TH-2021-017, PSI-PR-21-02, ZU-TH 04/21",
    doi = "10.1103/PhysRevLett.127.071801",
    journal = "Phys. Rev. Lett.",
    volume = "127",
    number = "7",
    pages = "071801",
    year = "2021"
}

@article{Escribano:2023seb,
    author = "Escribano, Rafel and Miranda, Jes{\'u}s Alejandro and Roig, Pablo",
    title = "{Radiative corrections to the {\ensuremath{\tau}}-{\textrightarrow}(P1P2){\ensuremath{-}}v{\ensuremath{\tau}} (P1,2={\ensuremath{\pi}},K) decays}",
    eprint = "2303.01362",
    archivePrefix = "arXiv",
    primaryClass = "hep-ph",
    doi = "10.1103/PhysRevD.109.053003",
    journal = "Phys. Rev. D",
    volume = "109",
    number = "5",
    pages = "053003",
    year = "2024"
}

@article{Antonelli:2013usa,
    author = "Antonelli, Mario and Cirigliano, Vincenzo and Lusiani, Alberto and Passemar, Emilie",
    title = "{Predicting the $\tau$ strange branching ratios and implications for $V_{us}$}",
    eprint = "1304.8134",
    archivePrefix = "arXiv",
    primaryClass = "hep-ph",
    reportNumber = "LA-UR-13-22949",
    doi = "10.1007/JHEP10(2013)070",
    journal = "JHEP",
    volume = "10",
    pages = "070",
    year = "2013"
}

@article{Flores-Baez:2013eba,
    author = "Flores-Ba{\'e}z, F. V. and Morones-Ibarra, J. R.",
    title = "{Model Independent Electromagnetic corrections in hadronic $\tau$ decays}",
    eprint = "1307.1912",
    archivePrefix = "arXiv",
    primaryClass = "hep-ph",
    doi = "10.1103/PhysRevD.88.073009",
    journal = "Phys. Rev. D",
    volume = "88",
    number = "7",
    pages = "073009",
    year = "2013"
}

@article{Cirigliano:2008wn,
    author = "Cirigliano, Vincenzo and Giannotti, Maurizio and Neufeld, Helmut",
    title = "{Electromagnetic effects in K(l3) decays}",
    eprint = "0807.4507",
    archivePrefix = "arXiv",
    primaryClass = "hep-ph",
    reportNumber = "LA-UR-07-7940, UWTHPH-2008-12",
    doi = "10.1088/1126-6708/2008/11/006",
    journal = "JHEP",
    volume = "11",
    pages = "006",
    year = "2008"
}

\end{document}